\documentclass[twocolumn,preprintnumbers,amsmath,amssymb]{revtex4}  

\usepackage{amsfonts}
\usepackage[T1]{fontenc}
\usepackage{amsmath,amsbsy,amssymb,graphicx}
\usepackage{times}
\usepackage{amsmath}
\usepackage{amssymb}
\usepackage{graphicx}
\usepackage{color}

\begin{document}

\title{Valley-contrasting interband transitions and excitons in symmetrically biased dice model}

\author{Lei Hao}
 \address{School of Physics, Southeast University, Nanjing 211189, China}

\date{\today}

\begin{abstract}
We study the exciton states in the symmetrically biased dice model, the electronic structures of which have an isolated flat band between two dispersive bands. At 1/3 or 2/3 filling, the model describes a two-dimensional semiconductor with the band edge at two degenerate valleys. Because of qualitative changes in the eigenvectors resulting from the bias term, the interband transition between the flat band and a dispersive band is valley contrasting under circularly polarized light. In terms of an effective-mass model and a realistic electron-hole interaction, we numerically calculate the spectrum and wave functions of the intravalley excitons, which are treated as Wannier-Mott excitons. We also discuss the fine structures of the exciton spectrum induced by the intravalley and intervalley exchange interactions. The symmetrically biased dice model thereby proves to be a new platform for valley-contrasting optoelectronics.
\end{abstract}


\maketitle

\section{introduction}

Two-dimensional (2D) electronic systems with flat bands feature a plethora of intriguing single-particle properties and many-body phases.
Among others, these include the super-Klein (or, all-angle Klein) tunneling \cite{shen10,urban11,illes17,betancur17}, the high Chern number quantum anomalous Hall states \cite{wang11,andrijauskas15,su19}, novel quantum Hall series \cite{lan11,illes15,illes16,biswas16,xu17}, prone to itinerant ferromagnetism at weak interaction strength \cite{lieb89,zhang10}, possible high-temperature superconductivity \cite{kopnin11,lothman17}, and so on.

The two-body or few-body bound states, such as the excitons or trions in 2D semiconductors and insulators, are another important kind of states. Because of the reduced dielectric screening in 2D systems, the exciton binding energies of 2D semiconductors are usually much larger than those in 3D systems \cite{cudazzo10,cudazzo11,latini15}. The excitons therefore have profound influence on the optoelectronic properties of 2D semiconductors and insulators \cite{ramasubramaniam12,qiu13,ugeda14,he14,bernardi13,jariwala14,lopez13,baugher14,ross14,pospischil14,berkelbach13,wu15,qiu15,pulci12,chernikov14,walther18}. In addition, the high exciton binding energy also makes 2D semiconductors more promising than 3D semiconductors to become excitonic insulators \cite{du17,wang19,varsano20}.
Excitons in 2D systems with flat bands were considered theoretically in the case where electrons and holes are both inside completely flat bands \cite{ishii02,ishii04}. The excitons were found to have remarkably large binding energies and are extremely localized small (Frenkel) excitons \cite{ishii02}. Another theory studied the exciton binding energy in systems with a nearly flat band as the valence band and a dispersive conduction band, concluding that the flatness of the valence band reduces the exciton binding energy \cite{trushin19}. The excitons were treated as Wannier-Mott excitons in the latter work. Recently, there have been several experiments on the exciton effects in systems with flat bands \cite{baboux16,ravets18,whittaker18}.

In this work, we study the excitons in a special 2D semiconductor with an isolated flat band and neighboring dispersive bands. Specifically, we consider the symmetrically biased dice model, in which a bias term breaks the inversion symmetry of the dice model and opens an energy gap between the flat band and the dispersive bands \cite{xu17,betancur17}. The band extrema of the model are located at two corners ($\mathbf{K}$ and $\mathbf{K}'$) of the Brillouin zone (BZ), which define a valley pseudospin $1/2$. We study the interband transitions between the flat band and a dispersive band, which we find to be valley contrasting under circularly polarized light, similar to the valley-contrasting optical transitions in transition metal dichalcogenides monolayers \cite{xiao12,cao12,jones13,yu14,sie15,wang18} and gapped graphene \cite{yao08}.

The valley-contrasting interband transitions suggest the symmetrically biased dice model to be another model system to study valley-contrasting optoelectronics. We thereby study the spectrum and wave functions of the excitons, considering a realistic attractive electron-hole interaction for 2D systems. By calculating the dielectric functions in the random-phase approximation, we show that the attractive electron-hole interaction is extended compared to the lattice parameter for weakly biased dice model. This implies that the excitons bound by the attractive interaction are also extended and may be considered as large Wannier-Mott excitons and studied in terms of a proper effective-mass model. Because the Wannier-Mott excitons are localized in the momentum space on one hand, and the interband transitions are valley contrasting on the other, we study the two valleys separately and calculate numerically the spectrum and relative wave functions of the intravalley excitons. For realistic parameters, the exciton levels are all within the semiconducting gap. The qualitative effects of the exchange interaction to the exciton spectrum are also discussed.

The remaining part of the paper is organized as follows. In Sec. II we describe the symmetrically biased dice model, show the valley-contrasting interband transitions, and introduce the formalism for calculating the spectrum and wave functions of the intravalley excitons. In Sec. III we present our results and discussions for the properties of intravalley excitons. Then we discuss in Sec. IV the qualitative effects of the exchange interaction, including a numerical calculation of the variation of the intravalley exciton spectrum with the intravalley exchange interaction. Finally, the main results are summarized in Sec. V.

\section{Model and method}

\subsection{Model and band structures}

As shown in Fig. 1(a), the dice lattice is a 2D lattice with three sublattices A, B, and C \cite{sutherland86,vidal98,rizzi06,bercioux09}. The nonvanishing hopping integrals consist of those connecting A or C sites with nearest-neighboring (NN) B sites. The low-energy band structures of electrons in these lattices are described by the following tight-binding model
\begin{eqnarray}
\hat{H}&=&\sum\limits_{\langle i,j\rangle,\sigma}(t_{ba}b^{\dagger}_{i\sigma}a_{j\sigma}
+t_{bc}b^{\dagger}_{i\sigma}c_{j\sigma}+\text{H.c.})   \notag \\
&&+\sum\limits_{i,\sigma}(\varepsilon_{ab}a^{\dagger}_{i\sigma}a_{i\sigma}
+\varepsilon_{cb}c^{\dagger}_{i\sigma}c_{i\sigma}).
\end{eqnarray}
The summation $\langle i,j\rangle$ runs over NN intersublattice sites. The index $\sigma$ labels the two spin components of the electrons. $a_{i\sigma}$, $b_{i\sigma}$, and $c_{i\sigma}$ separately annihilates a $\sigma$-spin electron on the A, B, and C sublattice of the $i$-th unit cell. H.c. means the Hermitian conjugate of the terms explicitly written out. We have taken the on-site energies for the B sublattice sites as reference, so $\varepsilon_{ab}=\varepsilon_{a}-\varepsilon_{b}$ and $\varepsilon_{cb}=\varepsilon_{c}-\varepsilon_{b}$. The three vectors connecting NN sites of two different sublattices include $\boldsymbol{\delta}_{1}=(-1,0)a_{0}$, $\boldsymbol{\delta}_{2}=(\frac{1}{2},-\frac{\sqrt{3}}{2})a_{0}$, and $\boldsymbol{\delta}_{3}=(\frac{1}{2},\frac{\sqrt{3}}{2})a_{0}$. The two primitive lattice vectors are $\mathbf{a}_{1}=\boldsymbol{\delta}_{2}-\boldsymbol{\delta}_{1}=(\frac{\sqrt{3}}{2},-\frac{1}{2})a$ and $\mathbf{a}_{2}=\boldsymbol{\delta}_{3}-\boldsymbol{\delta}_{1}=(\frac{\sqrt{3}}{2},\frac{1}{2})a$, where $a=\sqrt{3}a_{0}$.

\begin{figure}[!htb]\label{fig1} \centering
\hspace{-2.95cm} {\textbf{(a)}} \hspace{3.8cm}{\textbf{(b)}}\\
\hspace{0cm}\includegraphics[width=4.5cm,height=3.5cm]{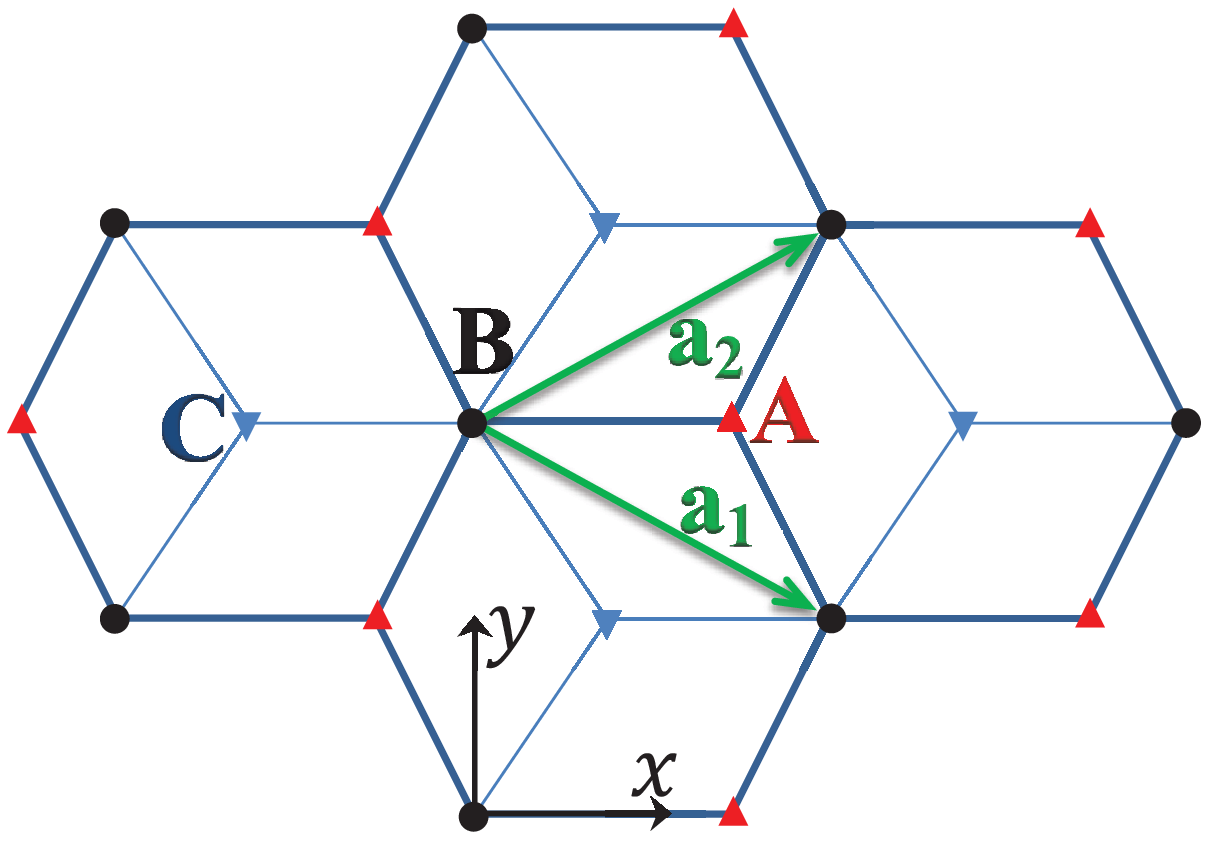}
\includegraphics[width=3.68cm,height=3.3cm]{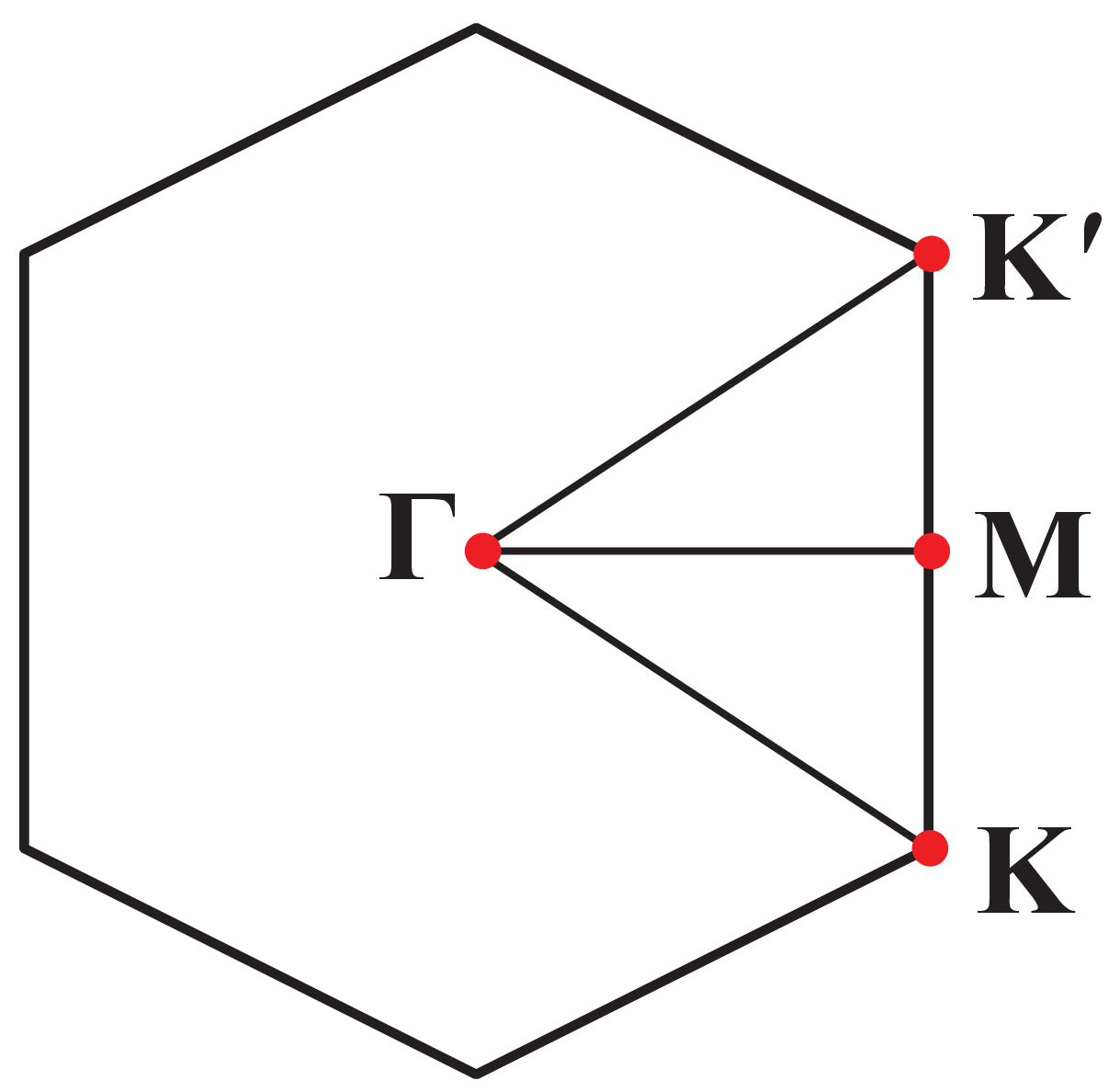}\\
\hspace{-7.2cm} {\textbf{(c)}} \\
\hspace{0cm}\includegraphics[width=8cm,height=5.75cm]{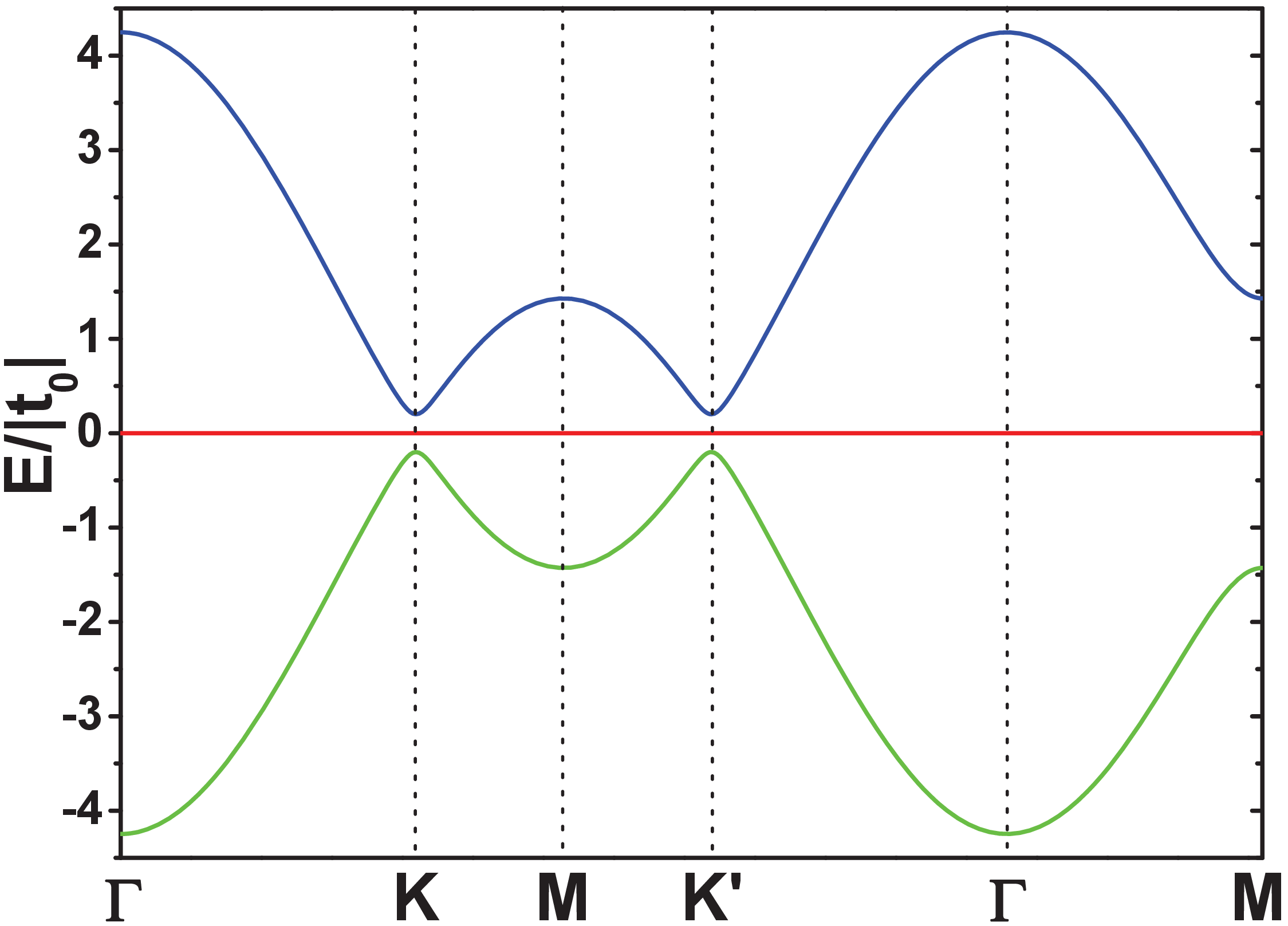} \\
\vspace{-0.10cm}
\caption{(a) The dice lattice. (b) Brillouin zone (BZ) of the dice lattice. The high symmetry points of the BZ are indicated. (c) The band structures for the parameters $|t_{0}|=1$ and $|\Delta|=0.2|t_{0}|$, along several high-symmetry directions of the BZ.}
\end{figure}

We consider the parameter combinations $t_{ba}=t_{bc}=t_{0}$ and $\varepsilon_{ab}=-\varepsilon_{cb}=\Delta$, which defines a symmetrically biased dice model \cite{betancur17,xu17}. Making Fourier transformation to the reciprocal space, and introducing the basis $\psi^{\dagger}_{\mathbf{k}\sigma}=[a^{\dagger}_{\mathbf{k}\sigma},b^{\dagger}_{\mathbf{k}\sigma},c^{\dagger}_{\mathbf{k}\sigma}]$, we have
\begin{equation}
\hat{H}=\sum\limits_{\mathbf{k}\sigma}\psi^{\dagger}_{\mathbf{k}\sigma}h(\mathbf{k})\psi_{\mathbf{k}\sigma},
\end{equation}
where
\begin{equation}
h(\mathbf{k})=\begin{pmatrix} \Delta & \xi_{ab}(\mathbf{k}) & 0  \\
                              \xi_{ab}^{\ast}(\mathbf{k}) & 0 & \xi_{bc}(\mathbf{k}) \\
                              0 & \xi_{bc}^{\ast}(\mathbf{k}) & -\Delta
               \end{pmatrix}.
\end{equation}
The off-diagonal elements are
\begin{equation}
\xi(\mathbf{k})=\xi_{ab}(\mathbf{k})=\xi_{bc}(\mathbf{k}) =t_{0}(e^{i\mathbf{k}\cdot\boldsymbol{\delta}_{1}}+e^{i\mathbf{k}\cdot\boldsymbol{\delta}_{2}}+e^{i\mathbf{k}\cdot\boldsymbol{\delta}_{3}}).
\end{equation}
For $\Delta\neq0$, the band structure consists of one flat band at $E_{0}(\mathbf{k})=0$ and two dispersive bands at
\begin{equation}
E_{\nu}(\mathbf{k})=\nu\sqrt{\Delta^{2}+2|\xi(\mathbf{k})|^{2}}\equiv\nu E(\mathbf{k}),
\end{equation}
where $\nu=\pm$. All three bands are two-fold degenerate by spin. The band structures for a typical set of parameters are shown in Fig. 1(c). At a filling fraction of 1/3 or 2/3 the system is in an insulating state. The band gap attains the minimum $|\Delta|$ at $\mathbf{K}=(\frac{\sqrt{3}}{2},-\frac{1}{2})\frac{4\pi}{3a}$ and $\mathbf{K}'=(\frac{\sqrt{3}}{2},\frac{1}{2})\frac{4\pi}{3a}$ of the BZ, where $\xi(\mathbf{k})=0$. The band edges thus situate at these two distinct $\mathbf{K}$ points, which define a two-fold degenerate valley degree of freedom. We will focus on the $\frac{1}{3}$-filled systems, in which the flat band is empty and the impact of the on-site correlation may therefore be neglected more safely. We then focus on the long-range part of the Coulomb interaction responsible for the formation of excitons.

The (biased) dice model of fermions may be realized in several systems. Firstly, as was pointed out by Wang and Ran, three consecutive (111) layers of a cubic lattice constitutes the displaced dice lattice \cite{wang11}. The dice model is realized if a tight-binding model up to hoppings between NN pairs of sites provides an excellent description for the low-energy electronic structures. The symmetrically biased dice model is then obtained by applying a uniform electric field perpendicular to the layer planes. Secondly, we may imagine to fabricate a triangular lattice of quantum dots with the connections between the dots following the pattern of the dice lattice \cite{ansaloni20,sikdar21}. When all the quantum dots have the same energy level structures with a single orbit close to the chemical potential, the dice model is realized. By biasing the on-site energies of the A sublattice sites and the C sublattice sites oppositely, we get the symmetrically biased dice model. Thirdly, the dice model may be realized in optical lattices of fermionic cold atoms \cite{rizzi06}. In particular, both the pure dice model and the symmetrically biased dice model may be realized by fermionic cold atoms on a tunable displaced optical dice lattice \cite{hao20}. Since we are mainly interested in the electronic systems, we will keep the first and second systems in mind in the following discussions. In this work, we assume a negligible spin-orbit coupling in the system. A finite spin-orbit coupling may render the model topologically nontrivial \cite{wang11,wang21}.

\subsection{Valley-contrasting interband transitions}

Manipulating the valley degrees of freedom constitutes the basis of valleytronics \cite{schaibley16}. Several materials, including the gapped monolayer graphene \cite{yao08} and the monolayer transition metal dichalcogenides \cite{xiao12,cao12,jones13,yu14,sie15,wang18}, allow the optical implementation of valleytronics. These systems both have a band structure containing two valleys and lack the inversion symmetry. A circularly polarized light selectively excites interband transitions in one of the two valleys, while the oppositely polarized light excites the interband transitions in the other valley. The crucial importance of broken inversion symmetry to the valley-contrasting interband transitions was pointed out by Yao et al \cite{yao08}, who showed that the degree of circular polarization of the interband transition at a wave vector $\mathbf{k}$ in the BZ is proportional to the orbital magnetic moment $\mathbf{m}_{\text{orb}}(\mathbf{k})$ at $\mathbf{k}$. Since the orbital magnetic moment as an axial vector is symmetric with respect to the inversion operation, valley-contrasting interband transitions can occur only if the inversion symmetry is broken \cite{yao08}.

The symmetrically biased dice model, while having a completely flat band, also has broken inversion symmetry and two valleys. To see whether the symmetrically biased dice model has valley-contrasting interband transitions, we firstly write down the eigenvectors of the states in the three bands. Hereafter we neglect the dummy spin label.
Up to an arbitrary $U(1)$ phase factor, we take the eigenvectors for states of the flat band as
\begin{equation}
|\psi_{0}(\mathbf{k})\rangle=\frac{1}{E(\mathbf{k})}\begin{pmatrix} -\xi(\mathbf{k}) \\ \Delta \\ \xi^{\ast}(\mathbf{k}) \end{pmatrix}.
\end{equation}
For states of the band $E_{\nu}(\mathbf{k})$ ($\nu=\pm$) defined by Eq.(5), we choose the eigenvectors as
\begin{equation}
|\psi_{\nu}(\mathbf{k})\rangle=\frac{|\xi(\mathbf{k})|}{E(\mathbf{k})}
\begin{pmatrix} \frac{\xi(\mathbf{k})}{E(\mathbf{k})-\nu\Delta}  \\ \nu  \\ \frac{\xi^{\ast}(\mathbf{k})}{E(\mathbf{k})+\nu\Delta} \end{pmatrix}.
\end{equation}

The parameter $\Delta$ brings dramatic changes to the eigenvectors close to the $\mathbf{K}$ and $\mathbf{K}'$ points of the BZ.
To see this, we expand the eigenvectors into the polynomials of the relative wave vector $\mathbf{q}=\mathbf{k}-\mathbf{K}_{\tau}$ ($\tau=\pm$, $\mathbf{K}_{-}=\mathbf{K}$ and $\mathbf{K}_{+}=\mathbf{K}'$). We have
\begin{equation}
|\psi_{0}(\mathbf{K}_{\tau}+\mathbf{q})\rangle\simeq\frac{1}{|\Delta|}\begin{pmatrix} \frac{\sqrt{3}}{2}at_{0}e^{-i\frac{\pi}{6}}(q_{x}+i\tau q_{y})  \\ \Delta  \\ -\frac{\sqrt{3}}{2}at_{0}e^{i\frac{\pi}{6}}(q_{x}-i\tau q_{y}) \end{pmatrix},
\end{equation}
and
\begin{equation}
|\psi_{\nu}(\mathbf{K}_{\tau}+\mathbf{q})\rangle\simeq\frac{\frac{\sqrt{3}}{2}a|t_{0}|q}{|\Delta|}
\begin{pmatrix} \frac{-\frac{\sqrt{3}}{2}at_{0}e^{-i\frac{\pi}{6}}(q_{x}+i\tau q_{y})} {[1-\nu\text{sgn}(\Delta)]|\Delta|+\frac{3a^{2}t_{0}^{2}q^{2}}{4|\Delta|}} \\ \nu \\ \frac{-\frac{\sqrt{3}}{2}at_{0}e^{i\frac{\pi}{6}}(q_{x}-i\tau q_{y})} {[1+\nu\text{sgn}(\Delta)]|\Delta|+\frac{3a^{2}t_{0}^{2}q^{2}}{4|\Delta|}} \end{pmatrix}.
\end{equation}
$q=\sqrt{q_{x}^{2}+q_{y}^{2}}$.
Close to the $\mathbf{K}$ or $\mathbf{K}'$ points, where $q_{x},q_{y}\simeq0$, states of the flat band come mainly from the B sublattice. In contrast, as is clear from Eq.(6), the weight on the B sublattice is zero for states of the flat band at $\Delta=0$. This is the most dramatic change brought to the states of the flat band by a finite $\Delta$, although the band is still completely flat. For $\Delta=0$, the states of the dispersive band defined by Eq.(7) are linear combinations of the A, B, and C sublattices at the probability of $\frac{1}{4}$$:$$\frac{1}{2}$$:$$\frac{1}{4}$ for all $\mathbf{k}$. For $\Delta\neq0$, however, states in the dispersive bands close to the band edges consist only of states of the A or C sublattices. More specifically, for $\nu\text{sgn}(\Delta)=1$ [$\nu\text{sgn}(\Delta)=-1$], the state defined by Eq.(9) comes mainly from the A (C) sublattice.

We are interested in the resonant interband transitions close to the two valleys. The interband transition between the flat band and the dispersive bands may be induced by applying an optical field with a frequency resonant with the transition. For the $1/3$ or $2/3$ filled system, this corresponds to $\hbar\omega\simeq|\Delta|$. In the dipole approximation, the strength of the vertical interband transition is proportional to \cite{yao08,xiao12}
\begin{equation}
\langle\psi_{f}(\mathbf{k})|\mathbf{n}\cdot\boldsymbol{\nabla}_{\mathbf{k}}h(\mathbf{k})
|\psi_{i}(\mathbf{k})\rangle,
\end{equation}
where $|\psi_{i}(\mathbf{k})\rangle$ and $|\psi_{f}(\mathbf{k})\rangle$ are the initial and final states of the interband transition at $\mathbf{k}$. For $1/3$-filled bands, $\psi_{i}=\psi_{-}$ and $\psi_{f}=\psi_{0}$. For $2/3$-filled bands, $\psi_{i}=\psi_{0}$ and $\psi_{f}=\psi_{+}$. The unit vector $\mathbf{n}$ is the polarization vector (Jones vector) of the light. In particular, $\mathbf{n}=(1,i\eta)/\sqrt{2}$ with $\eta=\pm1$ corresponds to circularly polarized light. $\eta$ represents the chirality of the circularly polarized light. By substituting Eq.(3) and the eigenvectors Eqs.(6)-(9) into Eq.(10), and expanding the results close to the two valleys, we get, up to the leading order of $\mathbf{q}=\mathbf{k}-\mathbf{K}_{\tau}$,
\begin{eqnarray}
&&\langle\psi_{0}(\mathbf{k})|\frac{1}{\sqrt{2}}[\frac{\partial h(\mathbf{k})}{\partial k_{x}}+i\eta\frac{\partial h(\mathbf{k})}{\partial k_{y}}]
|\psi_{\nu}(\mathbf{k})\rangle  \notag \\
&&=\frac{\sqrt{3}a}{2\sqrt{2}}|t_{0}|[\text{sgn}(\Delta)+\nu\eta\tau]\frac{q_{x}+i\eta q_{y}}{q}.
\end{eqnarray}
This is one of the central results of the present work.
For states of the $\tau$-valley, the matrix element is nonzero only for circularly polarized light satisfying $\eta=\nu\tau\text{sgn}(\Delta)$. As the valley index $\tau$ changes sign, the chirality index $\eta$ changes its sign alongside. Therefore, the interband transition close to the band edges between the flat band and one dispersive band is valley contrasting. This valley-contrasting interband transition is a direct consequence of the changes brought to the eigenvectors by the parameter $\Delta$, as was explained above. In contrast, the current operator is unchanged by $\Delta$.
Consistent with this picture, the matrix elements between the flat band and the linearly dispersive bands are zero at the two valleys for $\Delta=0$.

We also note that, which may be verified directly, the matrix element of Eq.(10) vanishes between the two dispersive bands at the band edges. This is again consistent with the character of the eigenvectors of the two dispersive bands, since the current operator couples states of the A or C sublattices to states of the B sublattice whereas the states of the two dispersive bands consist separately of states of the A and C sublattices.

Overall, we have seen that the bias term ($\Delta\neq0$) leads to valley-contrasting interband transitions close to the band edges at $\mathbf{K}$ and $\mathbf{K}'$. This suggests that the symmetrically biased dice model is another very interesting model system that supports valley-contrasting optoelectronics. For this purpose, it is important to study the properties of exciton excitations in this system. We will focus in what follows on the excitons in systems at $1/3$ filling.

\subsection{Effective-mass model for intravalley excitons}

Apart from excitons in molecular crystals and excitons in ionic crystals with small dielectric constants (such as the alkali halides), excitons in most insulators and semiconductors may be treated as large Wannier-Mott excitons \cite{bassanibook,knoxbook}. The wave functions for the relative motion of the electron and hole states of the large excitons are very extended in the real space, covering a region much larger than the unit cell of the lattice. In the $\mathbf{k}$ space, correspondingly, the electron and hole states relevant to the formation of large excitons are highly concentrated and only consist of the states close to the extrema of the conduction and valence bands \cite{bassanibook,knoxbook}. The present studies of the excitons have two major objectives. Firstly, we show that the Wannier-Mott model is applicable to the excitons of the present system at weak bias (i.e., for small $|\Delta|$), despite the presence of the completely flat band. Secondly, we want to compare the band gap $|\Delta|$ and the binding energy of the lowest energy exciton states. When $|\Delta|$ is larger than the binding energy of the lowest energy exciton states we have a typical semiconductor. Otherwise, the semiconductor is unstable to the transition to an excitonic insulator.

The symmetrically biased dice model at $1/3$ filling has a flat band as the conduction band and a dispersive band as the valence band. The spatial extension of the exciton state should be a compromise between localized electron state of the flat conduction band and extended hole state of the dispersive valence band. Since the dispersive band becomes increasingly flat as $|\Delta|$ increases, we suppose the composite electron-hole pairs to be increasingly compact as $|\Delta|$ increases. Conversely, for small $|\Delta|$ the excitons may be sufficiently extended to allow for a description as large Wannier-Mott excitons. We firstly assume that it is indeed the case for the $1/3$-filled system. Criterion supporting this view is provided in the following discussions.

The Wannier-Mott excitons may be studied in terms of a low-energy effective-mass model defined in terms of the states in the neighborhood of the band extrema. In the present system, in view of the valley-contrasting interband excitations under circularly polarized light and the localization of states in the $\mathbf{k}$ space relevant to the exciton formation, the Wannier-Mott excitons may be studied by focusing on the states close to a single valley. We thus treat the two valleys independently and get two series of exciton modes distinguished by the valley index. This is similar to the procedure taken for monolayer transition-metal dichalcogenides \cite{jones13}. Besides the presence of a completely flat band in the present system, another difference lies in the negligence of spin-orbit coupling, in contrast to the strong spin-orbit coupling in monolayer transition-metal dichalcogenides \cite{xiao12}.

The hydrogenlike effective-mass model for the intravalley excitons of one valley is written as
\begin{equation}
\hat{H}=\hat{H}_{c}(\mathbf{p}_{e})+\hat{H}_{v}(\mathbf{p}_{h})-V_{\text{eff}}(\mathbf{r}_{e}-\mathbf{r}_{h}).
\end{equation}
$-V_{\text{eff}}(\mathbf{r}_{e}-\mathbf{r}_{h})$ is the effective electron-hole interaction, which contains both attractive long-range Coulomb interactions and the exchange interaction \cite{knoxbook,bassanibook}. For large Wannier-Mott excitons, we may retain only the long-range attractive Coulomb interaction in $-V_{\text{eff}}$ \cite{cudazzo11,bassanibook}. The exchange interaction produces fine structures of the exciton spectrum and will be discussed in Sec.IV. $\hat{H}_{c}(\mathbf{p}_{e})$ and $\hat{H}_{v}(\mathbf{p}_{h})$ are separately the effective-mass models for the electron state in the nearly empty conduction band and the hole state in the almost full valence band. Taking the top of the valence band $E_{-}(\mathbf{k})$ as the energy reference point, the dispersion of the electron band and the hole band are separately $E_{e}(\mathbf{k})=E_{0}(\mathbf{k})+|\Delta|$ and $E_{h}(\mathbf{k}')=-E_{-}(\mathbf{k}')-|\Delta|=E(\mathbf{k}')-|\Delta|$. $E_{0}(\mathbf{k})=0$, $E_{-}(\mathbf{k})$ and $E(\mathbf{k})$ are defined in Eq.(5). In the neighborhood of the band edge at $\mathbf{K}_{\tau}$ ($\tau=\pm$), we expand $E_{e}(\mathbf{k})$ and $E_{h}(\mathbf{k}')$ to the quadratic order of $\mathbf{k}-\mathbf{K}_{\tau}$ and $\mathbf{k}'-\mathbf{K}_{\tau}$. By defining $\mathbf{p}_{e}=\hbar(\mathbf{k}-\mathbf{K}_{\tau})$ and $\mathbf{p}_{h}=-\hbar(\mathbf{k}'-\mathbf{K}_{\tau})$, and taking them as the momentum operators for the motion of the electron and hole states close to $\mathbf{K}_{\tau}$, we obtain $\hat{H}_{c}(\mathbf{p}_{e})$ and $\hat{H}_{v}(\mathbf{p}_{h})$ from $E_{e}(\mathbf{k})$ and $E_{h}(\mathbf{k}')$ as follows
\begin{equation}
\begin{cases}
\hat{H}_{c}(\mathbf{p}_{e})=|\Delta|+\frac{\mathbf{p}_{e}^{2}}{2m_{e}},  \\
\hat{H}_{v}(\mathbf{p}_{h})=\frac{\mathbf{p}_{h}^{2}}{2m_{h}}, \\
\end{cases}
\end{equation}
where $m_{h}=2\hbar^{2}|\Delta|/(3t_{0}^{2}a^{2})$ is the effective mass of the hole states, $m_{e}=+\infty$ is the infinite mass of the electron states in the flat conduction band, $|\Delta|$ is the band gap. Eq.(13) is the same for both of the two valleys.

Introducing the center-of-mass (COM) coordinate $\mathbf{R}$ and the relative coordinate $\mathbf{r}$,
\begin{equation}
\begin{cases}
\mathbf{R}=\frac{m_{e}\mathbf{r}_{e}+m_{h}\mathbf{r}_{h}}{m_{e}+m_{h}}\simeq\mathbf{r}_{e},  \\
\mathbf{r}=\mathbf{r}_{e}-\mathbf{r}_{h}, \\
\end{cases}
\end{equation}
and the COM momentum and the relative momentum
\begin{equation}
\begin{cases}
\mathbf{P}=\mathbf{p}_{e}+\mathbf{p}_{h},  \\
\mathbf{p}=\frac{m_{h}\mathbf{p}_{e}-m_{e}\mathbf{p}_{h}}{m_{e}+m_{h}}\simeq-\mathbf{p}_{h}, \\
\end{cases}
\end{equation}
the full model is also decomposed into the part for the COM motion and the part for the relative motion
\begin{equation}
\begin{cases}
\hat{H}=\hat{H}_{\text{R}}+\hat{H}_{\text{r}},  \\
\hat{H}_{\text{R}}=\frac{\mathbf{P}^{2}}{2M^{\ast}} =-\frac{\hbar^{2}}{2M^{\ast}}\boldsymbol{\nabla}_{\mathbf{R}}^{2},  \\
\hat{H}_{\text{r}}=|\Delta|+\frac{\mathbf{p}^{2}}{2m^{\ast}}-V_{\text{eff}}(\mathbf{r}) =|\Delta|-\frac{\hbar^{2}}{2m^{\ast}}\boldsymbol{\nabla}_{\mathbf{r}}^{2}-V_{\text{eff}}(\mathbf{r}), \\
\end{cases}
\end{equation}
where the total mass for the COM motion $M^{\ast}=m_{e}+m_{h}\simeq m_{e}=\infty$, and the reduced mass of the relative motion $m^{\ast}=m_{e}m_{h}/(m_{e}+m_{h})\simeq m_{h}$.
Correspondingly, the wave function $\Psi(\mathbf{r}_{e},\mathbf{r}_{h})$ of the hydrogenlike electron-hole pair is decomposed into the part $\psi(\mathbf{R})$ for the COM motion and the part $\varphi(\mathbf{r})$ for the relative motion. The Coulomb interaction acts only on the relative motion and so the COM motion corresponds to a free particle. Therefore, $\psi(\mathbf{R})=A\exp(i\mathbf{k}_{\text{ex}}\cdot\mathbf{R})$. For finite COM momentum $\hbar\mathbf{k}_{\text{ex}}$, the kinetic energy for the COM motion is $\hbar^{2}\mathbf{k}_{\text{ex}}^{2}/(2M^{\ast})=0$. The excitons are therefore completely immobile as regards their COM motion. In other words, the intravalley exciton bands are completely flat. We therefore focus on the relative motion and take $\mathbf{k}_{\text{ex}}=0$. $\mathbf{k}_{\text{ex}}=0$ is consistent with excitons generated by light.

In $\mathbf{q}$-space, the screened long-range Coulomb interaction in an isotropic purely 2D system is known to be \cite{cudazzo11}
\begin{equation}
\phi_{\text{2D}}(\mathbf{q})=\frac{2\pi e^{2}}{\varepsilon(\mathbf{q})|\mathbf{q}|},
\end{equation}
where the 2D dielectric constant is $\mathbf{q}$-dependent and related to the 2D polarizability $\alpha_{\text{2D}}$ of the system via \begin{equation}
\varepsilon(\mathbf{q})=1+2\pi\alpha_{\text{2D}}|\mathbf{q}|.
\end{equation}
Fourier transformation of $\phi_{\text{2D}}(\mathbf{q})$ gives the real-space effective Coulomb interaction between two electrons or holes \cite{keldysh79,cudazzo11}
\begin{equation}
V_{\text{eff}}(\mathbf{r})=V_{\text{eff}}(r)=\frac{\pi e^{2}}{2r_{0}}[\text{H}_{0}(\frac{r}{r_{0}})-\text{Y}_{0}(\frac{r}{r_{0}})],
\end{equation}
where $r=|\mathbf{r}|$ is the length of $\mathbf{r}$, $r_{0}=2\pi\alpha_{\text{2D}}$. $\text{H}_{0}(x)$ is the zeroth order Struve function, and $\text{Y}_{0}(x)$ is the zeroth order second-kind Bessel function \cite{zhang96}. Eq.(19) was originally derived by Keldysh for thin semiconductor films \cite{keldysh79}. Recent analysis show that it provides a very accurate description for the effective screened Coulomb interaction for very thin quasi-2D systems \cite{cudazzo11,latini15}. We will take this form of interaction throughout the calculations.

The characteristic length $r_{0}=2\pi\alpha_{\text{2D}}$ defines the length scale of the effective Coulomb interaction $V_{\text{eff}}(r)$, which goes from an asymptotic logarithmic function for $r\ll r_{0}$ to the conventional $1/r$ behavior for $r\gg r_{0}$ \cite{cudazzo11}. Since the electron and hole states are bound by the effective Coulomb interaction into the exciton states, $r_{0}$ should also determine the spatial extension of the excitons. If $r_{0}$ is much larger than the lattice constant, it is a signal that treating the excitons as Wannier-Mott excitons is reasonable. It is therefore desirable to make an estimation over $r_{0}$ for the symmetrically biased dice model. This asks us to estimate $\alpha_{\text{2D}}$, which is the static polarizability of the electrons in the symmetrically biased dice model at $1/3$ filling. From Eq.(18), $\alpha_{\text{2D}}$ may be defined as the limit \cite{cudazzo11,latini15}
\begin{equation}
\alpha_{\text{2D}}=\text{lim}_{|\mathbf{q}|\rightarrow0}\frac{1}{|\mathbf{q}|}\frac{\varepsilon(\mathbf{q})-1}{2\pi}.
\end{equation}
The dielectric constant may be estimated in terms of the random phase approximation (RPA) by the following formula \cite{hwang07}
\begin{equation}
\varepsilon(\mathbf{q},\omega)=1-V(\mathbf{q})\chi_{0}(\mathbf{q},\omega),
\end{equation}
where the bare 2D Coulomb interaction $V(\mathbf{q})=2\pi e^{2}/(\varepsilon_{r}|\mathbf{q}|)$, the free polarizability $\chi_{0}(\mathbf{q},\omega)$ in the RPA corresponds to a bubble diagram \cite{hwang07}. Depending on the lattice structure and the model, the static dielectric constant evaluated by this approach may depend on the direction of $\mathbf{q}$. Correspondingly, instead of the $\mathbf{q}$-independent $\alpha_{\text{2D}}$ in Eq.(18), we may define a 2D polarizability which depends on the direction angle $\theta$ along which the wave vector $\mathbf{q}$ approaches zero
\begin{equation}
\alpha_{\text{2D}}(\theta)=-\text{lim}_{|\mathbf{q}|\rightarrow0}\frac{e^{2}}{\varepsilon_{r}|\mathbf{q}|^{2}}\chi_{0}(\mathbf{q},0),
\end{equation}
where $\mathbf{q}=|\mathbf{q}|(\cos\theta,\sin\theta)$. The relative dielectric constant $\varepsilon_{r}$ depends on the medium surrounding the 2D material. For graphene deposited on SiO$_{2}$, $\varepsilon_{r}=4$ was used in calculations \cite{hwang07}. For suspended 2D systems, it is reasonable to take $\varepsilon_{r}=1$. For simplicity, we will set $\varepsilon_{r}=1$ in what follows. A larger $\varepsilon_{r}$ does not change any qualitative conclusions. When the anisotropy in $\alpha_{\text{2D}}(\theta)$ is weak, we may approximate $\alpha_{\text{2D}}$ by $\alpha_{\text{2D}}(\theta)$ at a particular value of $\theta$ or by an average of $\alpha_{\text{2D}}(\theta)$ at its maximum and minimum \cite{rodin14}.

\begin{figure}[!htb]\label{fig2} \centering
\includegraphics[width=8.5cm,height=6.48cm]{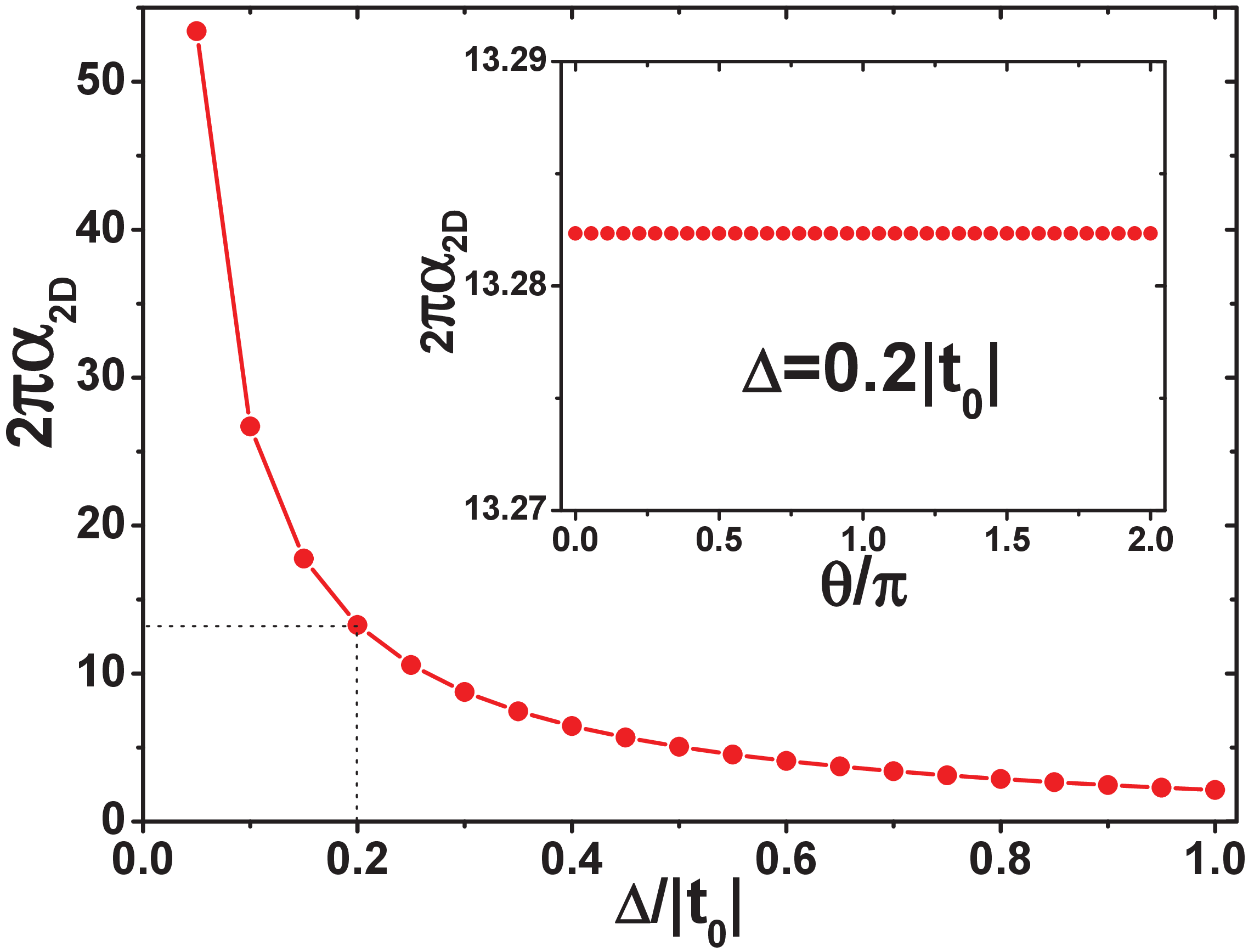}
\caption{The evolution of the 2D polarizability $\alpha_{\text{2D}}$ with $\Delta$, at fixed $t_{0}=-1$. The inset shows the evolution of $2\pi\alpha_{\text{2D}}$ with the angle $\theta$, defined in terms of Eq.(22), for $\Delta=0.2|t_{0}|$. $2\pi\alpha_{\text{2D}}=r_{0}$ is the length scale of the effective Coulomb interaction, which is in unit of $e^{2}/|t_{0}|=aE_{0}/|t_{0}|$. In other words, the vertical axis represents the ratio $\gamma$ between two lengths $r_{0}$ and $aE_{0}/|t_{0}|$.}
\end{figure}

In terms of Eqs.(20)-(22), we study the variation of the 2D polarizability $\alpha_{\text{2D}}$ with the parameter $\Delta$, at fixed filling fraction of $1/3$. We fix $|t_{0}|=-t_{0}=1$ as the energy unit and $a=1$ as the length unit.
For all the $\Delta$ studied, no discernible $\theta$-dependence is found in $\alpha_{\text{2D}}(\theta)$ defined by Eq.(22). The results for $\Delta=0.2|t_{0}|$ is shown as an example in the inset of Fig.2. The present model therefore gives an ideally isotropic $\alpha_{\text{2D}}$.
As shown in Fig. 2 is the evolution of $2\pi\alpha_{\text{2D}}=r_{0}$ with $\Delta$, in unit of
\begin{equation}
\frac{e^{2}}{a|t_{0}|}a=\frac{E_{0}}{|t_{0}|}a.
\end{equation}
Namely, Fig. 2 gives the variation of the ratio $\gamma$ between two lengths $r_{0}$ and $\frac{E_{0}}{|t_{0}|}a$. For $a=3$ {\AA}, the static electric energy (the formulae are in Gaussian units)
\begin{equation}
E_{0}=\frac{e^{2}}{a}\simeq 4.8 \hspace{0.1cm} \text{eV}.
\end{equation}
In comparison to the typical values of the hopping amplitudes in 2D solids, for example $|t_{0}|\simeq2.8$ eV for $a_{0}\simeq1.42$ {\AA} ($a=\sqrt{3}a_{0}\simeq2.46$ {\AA}) in graphene \cite{castroneto09}, we should have $E_{0}>|t_{0}|$ in general. With the increase of $a$, the hopping amplitude in a solid decays usually faster than $a^{-1}$ \cite{harrison80}. And so $E_{0}>|t_{0}|$ should survive the increase of $a$. If the dice lattice is realized as a metamaterial, such as a coupled array of quantum dots, the lattice parameters may be much larger than typical lattice constants of solids. In these metamaterials, the hopping amplitude may depend on the lattice constant in a manner different from that in solids. However, taking advantage of the high controllability in these artificial lattices, it should be able to satisfy the condition $E_{0}>|t_{0}|$. Then, if the value of $2\pi\alpha_{\text{2D}}$ (i.e., $\gamma$) in Fig. 2 is appreciably larger than 1, so that $r_{0}$ is much larger than $a$, it is reasonable to treat the excitons as Wannier-Mott excitons. From Fig. 2, $\gamma\simeq13.3$ for $\Delta=0.2|t_{0}|$. It therefore seems justified to treat the excitons in the symmetrically biased dice model for $\Delta$ close to $0.2|t_{0}|$ or smaller as Wannier-Mott excitons.
In numerical calculations for the energy spectrum and the wave functions of the excitons, we take $r_{0}=\beta a$ which amounts to $E_{0}=\beta|t_{0}|/\gamma$. From the above discussions we should usually have $\beta>\gamma$. Setting the lattice constant $a$ as the length unit, we have
\begin{equation}
\begin{cases}
\frac{\hbar^{2}}{2m^{\ast}a^{2}}=\frac{3t^{2}_{0}}{4|\Delta|}, \\
\frac{e^{2}}{r_{0}}=\frac{|t_{0}|}{\gamma}. \\
\end{cases}
\end{equation}
Now, $r_{0}=\beta a$ appears explicitly only in the Struve function and the second-kind Bessel function of Eq.(19).

We are now justified to calculate the spectrum and wave functions of the intravalley excitons by solving the eigenproblem of $\hat{H}_{\text{r}}$.
Because $\hat{H}_{\text{r}}$ has the rotational symmetry in the 2D plane of the dice lattice, we work in the 2D polar coordinate to study the relative motion of the excitons. In the polar coordinate $(\rho,\theta)$, $\mathbf{r}=\rho(\cos\theta,\sin\theta)$. Because the potential is independent of the angle $\theta$, the wave function $\varphi(\mathbf{r})$ for the relative motion should be the eigenstate of the angular momentum operator $l_{z}=-i\hbar\frac{\partial}{\partial\theta}$. We therefore take
\begin{equation}
\varphi(\mathbf{r})=\varphi(\rho,\theta)=\varphi_{nm}(\rho)\frac{e^{im\theta}}{\sqrt{2\pi}},
\end{equation}
where $n=1,2,3,...$ is the principal quantum number and $m=0,\pm1,\pm2,...$ is the angular quantum number characterizing the quantized $l_{z}$.
The differential equation for the radial wave function of the relative motion is
\begin{equation}
\{|\Delta|-\frac{\hbar^{2}}{2m^{\ast}}[\frac{d^{2}}{d\rho^{2}}+\frac{1}{\rho}\frac{d}{d\rho} -\frac{m^{2}}{\rho^{2}}]-V_{\text{eff}}(\rho)\}\varphi_{nm}(\rho)=E_{nm}\varphi_{nm}(\rho).
\end{equation}
For $m\neq0$, the above equation is invariant under $m\rightarrow -m$, all excitonic states with $m\neq0$ are therefore at least doubly degenerate. Only the excitons for $m=0$ can be optically active bright excitons and contribute to optical absorption, whereas all other exciton states are dark excitons \cite{bassanibook}. $E_{nm}$ increases with $n$ for fixed $m$. For each $m$, $n$ takes values among $|m|+1$, $|m|+2$, $\ldots$.

No exact analytical solution to Eq.(27) is known, because of the complexity of the potential $V_{\text{eff}}(\rho)$ defined by Eq.(19). Solutions were found in previous works by taking proper variational ansatz \cite{cudazzo11,pulci12,berkelbach13,chernikov14}, or purely numerical approaches \cite{walther18,chaves15,prada15}. Here, we follow Cudazzo \emph{et al} and expand the solutions to Eq.(27) into linear combinations of the eigenfunctions of the corresponding 2D hydrogen problem \cite{cudazzo11,yang91}. For this purpose, note that the effective Coulomb interaction of Eq.(19) becomes in the limit of large radius (i.e., $\rho\gg r_{0}$) \cite{zhang96}
\begin{equation}
V_{\text{eff}}(\rho)|_{\rho\gg r_{0}}\simeq\frac{e^{2}}{\rho}=V_{C}(\rho),
\end{equation}
which is the Coulomb interaction between two unit charges. By substituting $V_{C}(\rho)$ for $V_{\text{eff}}(\rho)$, Eq.(27) becomes the differential equation for the radial wave function of a 2D hydrogen, the exact analytical solutions of which give its spectrum $E^{(0)}_{nm}$ and the corresponding normalized radial wave functions $u_{nm}(\rho)$ \cite{yang91,zhang96}. Taking $u_{nm}(\rho)$ as the basis set for the unknown solutions to Eq.(27), we have
\begin{equation}
\varphi_{nm}(\rho)=\sum\limits_{n'}a^{(m)}_{nn'}u_{n'm}(\rho).
\end{equation}
The quantum number $n'=|m|+1,|m|+2,...$ runs over all the compatible states, the number of which is infinite. In practical numerical calculations, we truncate the summation over $n'$ up to the $N$ lowest values. This turns Eq.(27) into the eigenproblem of an $N\times N$ Hermitian matrix, which is the sum of a diagonal matrix with diagonal elements $E^{(0)}_{nm}$ ($n=|m|+1,...,|m|+N$) and a matrix whose elements are the matrix elements of
\begin{equation}
\delta V(\rho)=-[V_{\text{eff}}(\rho)-V_{C}(\rho)]
\end{equation}
with respect to $u_{nm}(\rho)$. That is
\begin{equation}
\delta V^{(m)}_{n_{1}n_{2}}=\int u_{n_{1}m}^{\ast}(\rho) \delta V(\rho) u_{n_{2}m}(\rho)\rho d\rho.
\end{equation}
The integration over $\rho$ ranges from 0 to $+\infty$. In practical calculations, we restrict the integration within a finite region $\rho\in[0,L]$, with $L\gg r_{0}$.

The solutions of the above $N\times N$ eigenproblem are approximations to $\varphi_{nm}(\rho)$. We then increase $N$ and test the convergence of the lowest several eigenvalues, until the changes in them are all smaller than a preset precision (e.g., smaller than $10^{-7}|t_{0}|$). $L$ is chosen so that $\varphi_{nm}(L)$ is vanishingly small for the lowest several eigenstates. The convergent solutions are reliable approximations to the lowest several eigenstates under consideration.

With the relative wave function calculated in the above manner, the total wave function of the excitons are obtained by multiplying it with the Bloch wave function $\psi_{0}(\mathbf{r}_{e})$ of the electron in the flat conduction band and the Bloch wave function $\psi_{-}(\mathbf{r}_{h})$ of the hole in the dispersive valence band,
\begin{equation}
\Psi_{\text{ex}}^{(n,m)}(\mathbf{r}_{e},\mathbf{r}_{h})=\varphi_{nm}(\rho)\frac{e^{im\theta}}{\sqrt{2\pi}} \psi_{0}(\mathbf{r}_{e})\mathcal{T}\psi_{-}(\mathbf{r}_{h}).
\end{equation}
$\rho$ and $\theta$ are separately the modulus and polar angle of $\mathbf{r}=\mathbf{r}_{e}-\mathbf{r}_{h}$. $\mathcal{T}$ is the time-reversal operation. In the effective-mass method, the Bloch wave functions are evaluated at the valley (i.e., at $\mathbf{K}_{\tau}$). The eigenvectors defined by Eq.(7) and Eq.(9) are however not well defined at $\mathbf{K}$ and $\mathbf{K}'$. This singularity is easily removed by multiplying the following $\nu$-dependent phase factor to Eqs.(7) and (9),
\begin{equation}
\frac{1+\nu\text{sgn}(\Delta)}{2}\frac{\xi^{\ast}(\mathbf{k})}{|\xi(\mathbf{k})|} +\frac{1-\nu\text{sgn}(\Delta)}{2}\frac{\xi(\mathbf{k})}{|\xi(\mathbf{k})|}.
\end{equation}
The same phase factor is to be multiplied to Eq.(11) after making this modification. No qualitative conclusion is changed by this phase factor.

Since the present model [i.e., Eq.(1)] is independent of spin, each exciton level has a four-fold degeneracy in the total spin, one for the singlet channel and three for the triplet channel \cite{bassanibook}.

\section{Properties of intravalley excitons}

For the symmetrically biased dice model, only the states close to the $\mathbf{K}_{\tau}$ ($\tau=\pm$) points are relevant to the formation of large Wannier-Mott excitons. Since these two valleys are well separated, it is a good approximation to focus on a single valley in the first place and taking the influence of the intervalley interactions into consideration afterwards. In particular, because of the valley-contrasting interband transitions, the intravalley excitons are relevant to the excitons generated in experiments by a circularly polarized light. We therefore study the spectrum and wave function of intravalley excitons in terms of the effective-mass model defined in the previous section, which is applicable to both of the two valleys.

\begin{figure}[!htb]\label{fig3} \centering
\hspace{-6.5cm} {\textbf{(a)}}\\
\includegraphics[width=8.0cm,height=5.88cm]{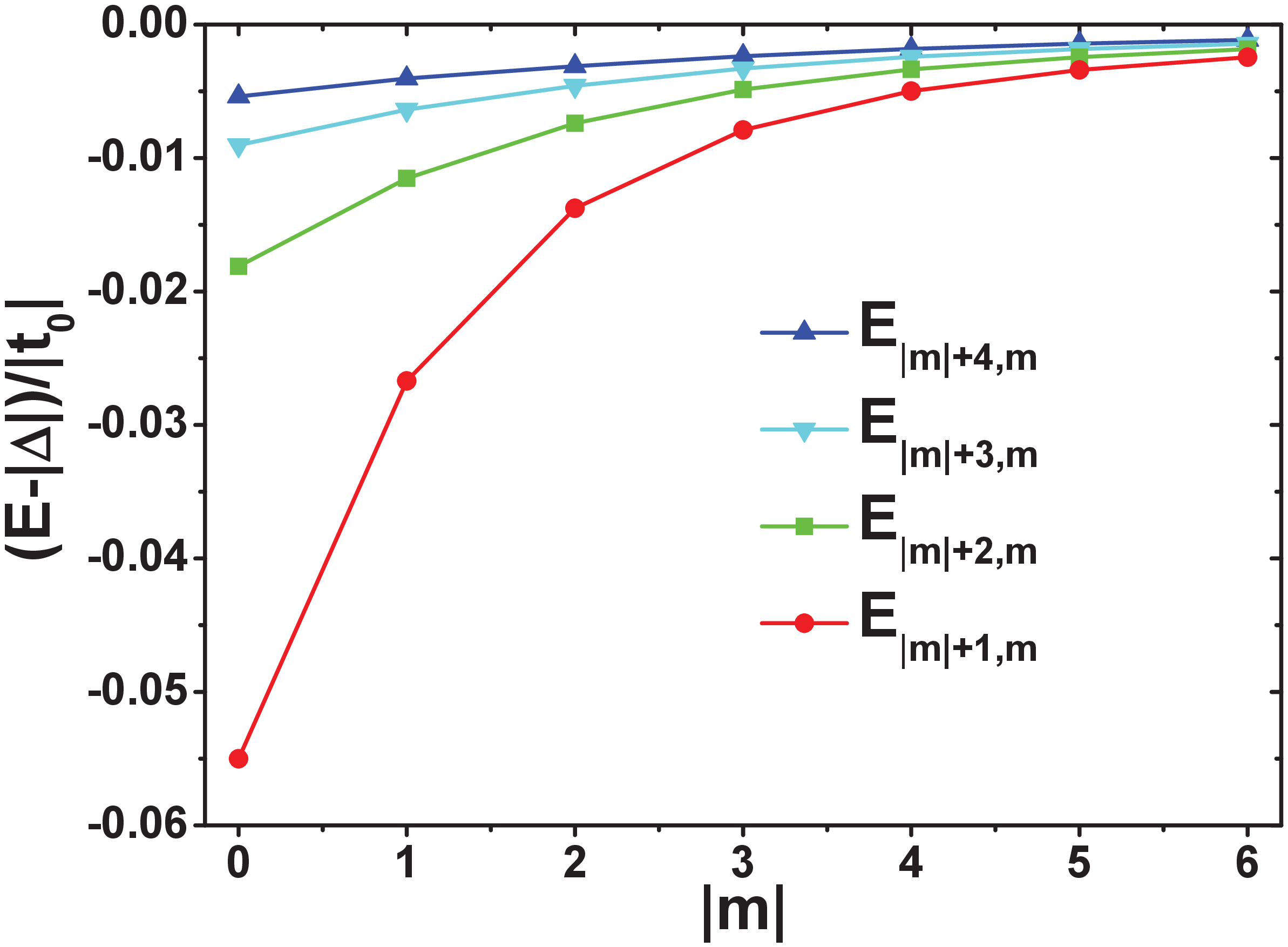} \\ \vspace{-0.05cm}
\hspace{-6.5cm} {\textbf{(b)}}\\
\includegraphics[width=8.0cm,height=5.81cm]{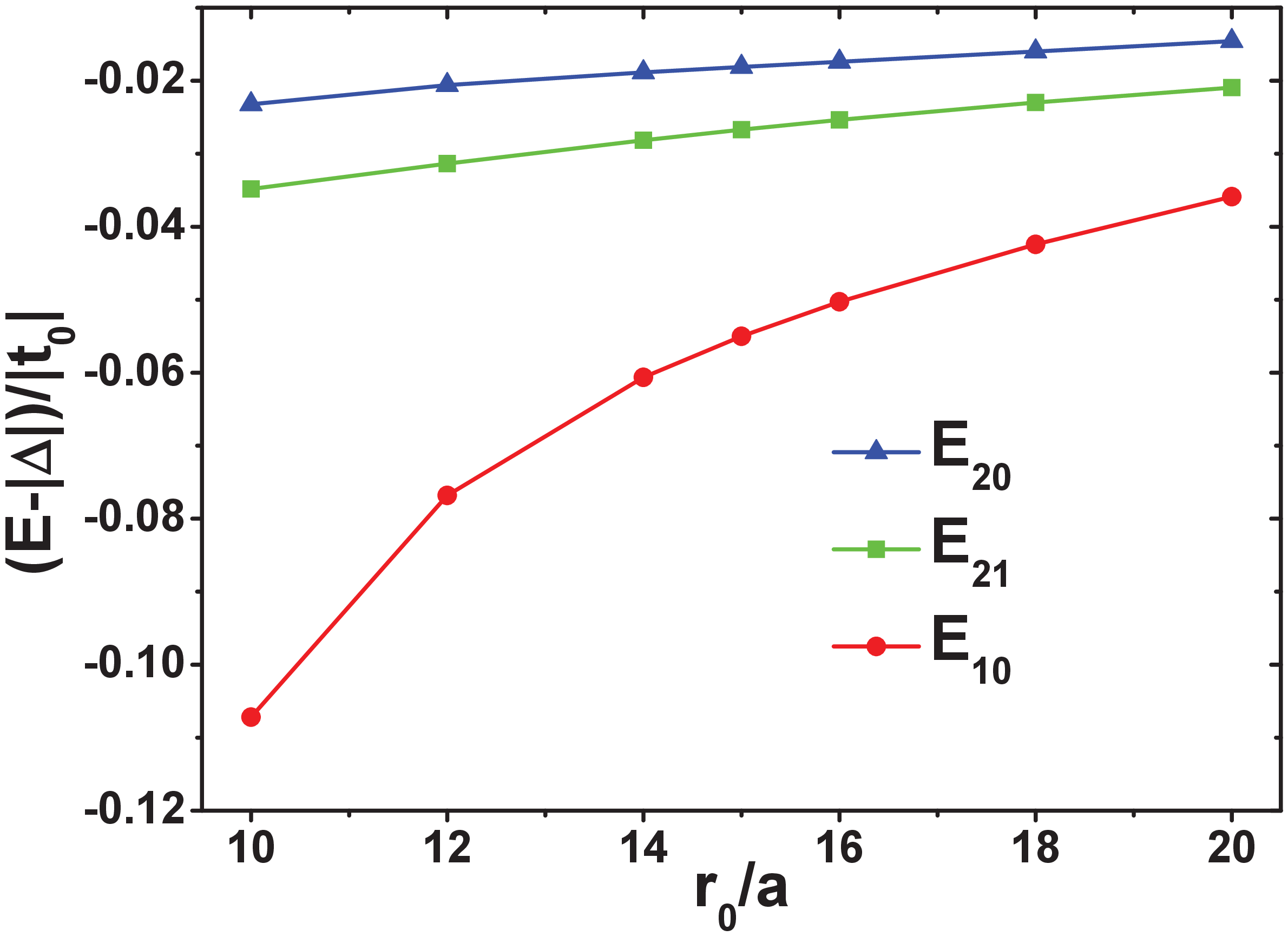}  \\
\caption{(a) The four lowest intravalley exciton modes, for $|m|$ ranging from 0 to 6. The energies are measured relative to the band gap, in unit of $|t_{0}|$. $t_{0}=-1$, $\Delta=0.2|t_{0}|$, and $r_{0}=\beta a=15a$. (b) Variations of the three lowest exciton modes with $r_{0}/a=\beta$.}
\end{figure}

In  terms of the formulae of Sec. IIC, we have calculated leading eigenstates of the intravalley excitons for several $m$, for $\Delta=0.2|t_{0}|$. We first take $r_{0}=15a$ ($\beta=15>13.3$). For $|m|$ ranging from 0 up to 6, the four lowest exciton states are shown in Fig. 3(a). The band gap $|\Delta|$ is subtracted from the exciton levels, so that the negative of the results are the corresponding binding energies. Convergence of the spectrum with respect to $N$ and $L$ are tested to ensure a precision in each of the five lowest eigenenergies to be smaller than $10^{-7}|t_{0}|$. Because the magnitude of the energy gap is $0.2|t_{0}|$, all the exciton levels are within the gap. The four exciton levels increase monotonically with $|m|$, so that the binding energies of the exciton levels decrease monotonically with $|m|$.

To see the dependence of the exciton spectrum on the parameter $\beta=r_{0}/a$, we study the variation of the three lowest exciton modes with $\beta$ from $\beta=10$ to $\beta=20$. All the other parameters are the same as those for Fig. 3(a). Together with the results for $\beta=15$, the variation of the spectrum are plotted in Fig. 3(b). The exciton levels (the binding energies) increase (decrease) monotonically with $r_{0}$. Down to $r_{0}=10a$, the exciton levels are all within the energy gap. It is therefore reasonable to take $\beta=15$ to see the qualitative properties of the excitons.

\begin{figure}[!htb]\label{fig4}
\centering
\hspace{-6.5cm} {\textbf{(a)}}\\
\includegraphics[width=8.0cm,height=5.56cm]{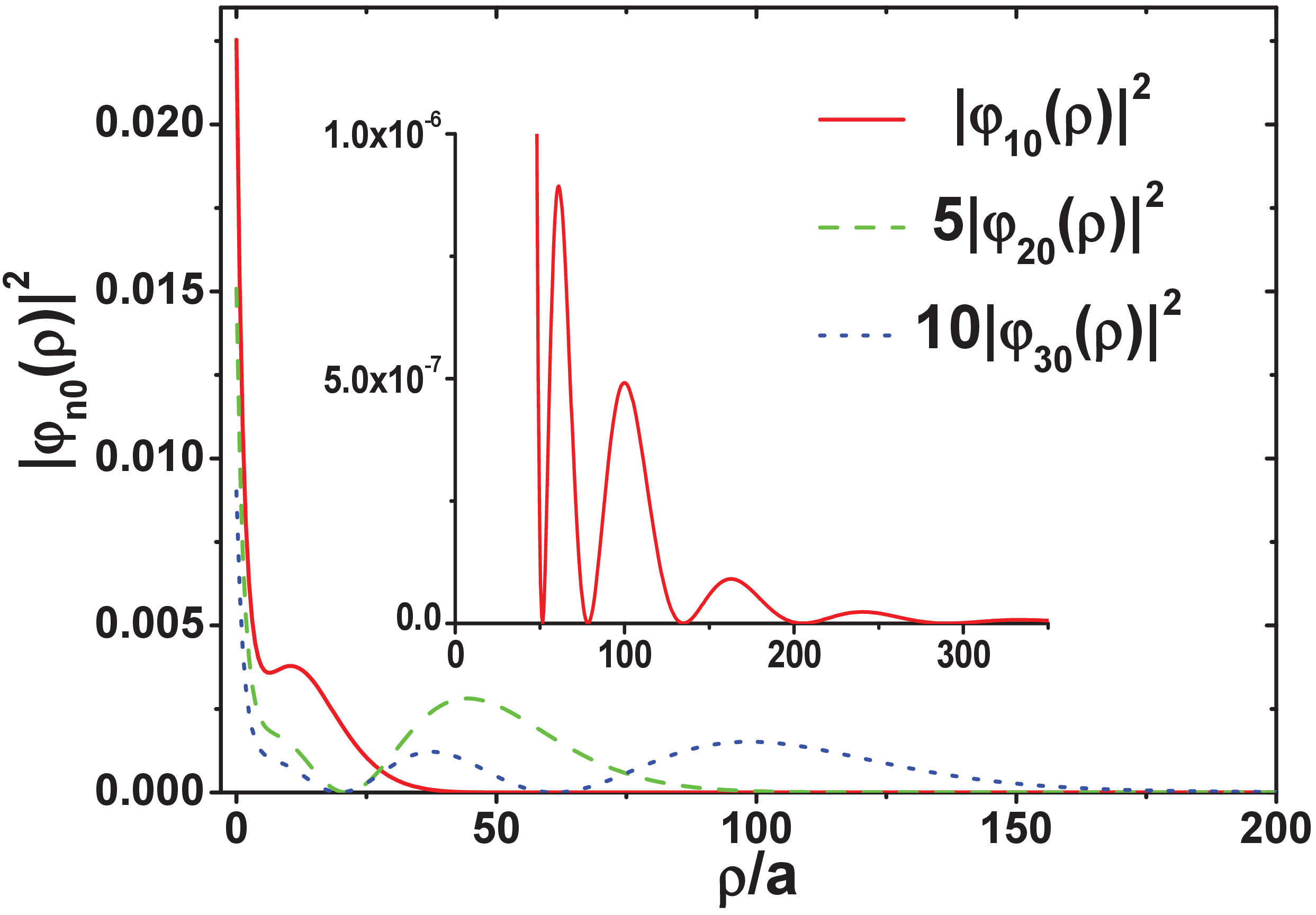} \\ \vspace{-0.05cm}
\hspace{-6.5cm} {\textbf{(b)}}\\
\includegraphics[width=8.0cm,height=5.42cm]{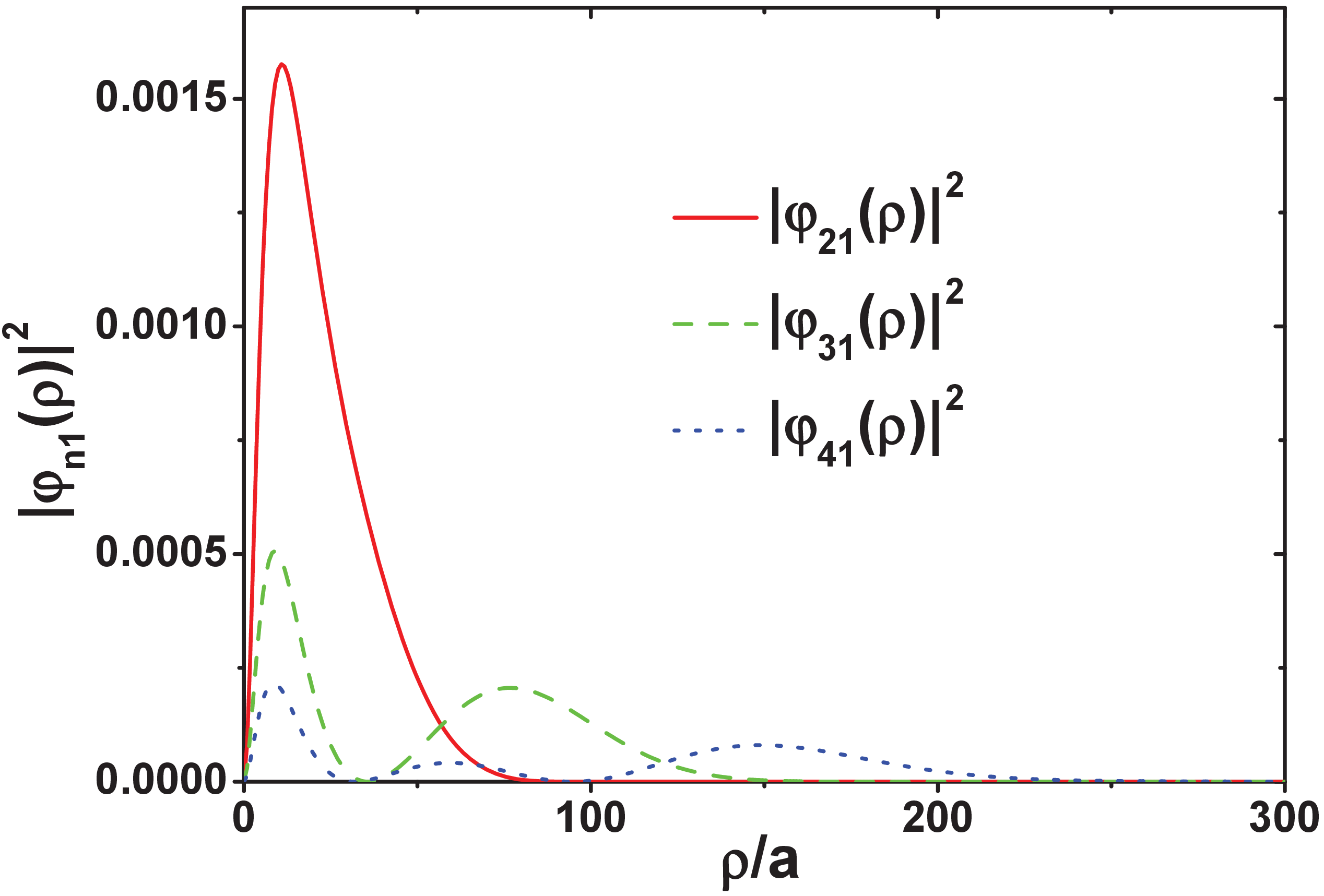}  \\ \vspace{-0.05cm}
\hspace{-6.5cm} {\textbf{(c)}}\\
\includegraphics[width=8.0cm,height=5.54cm]{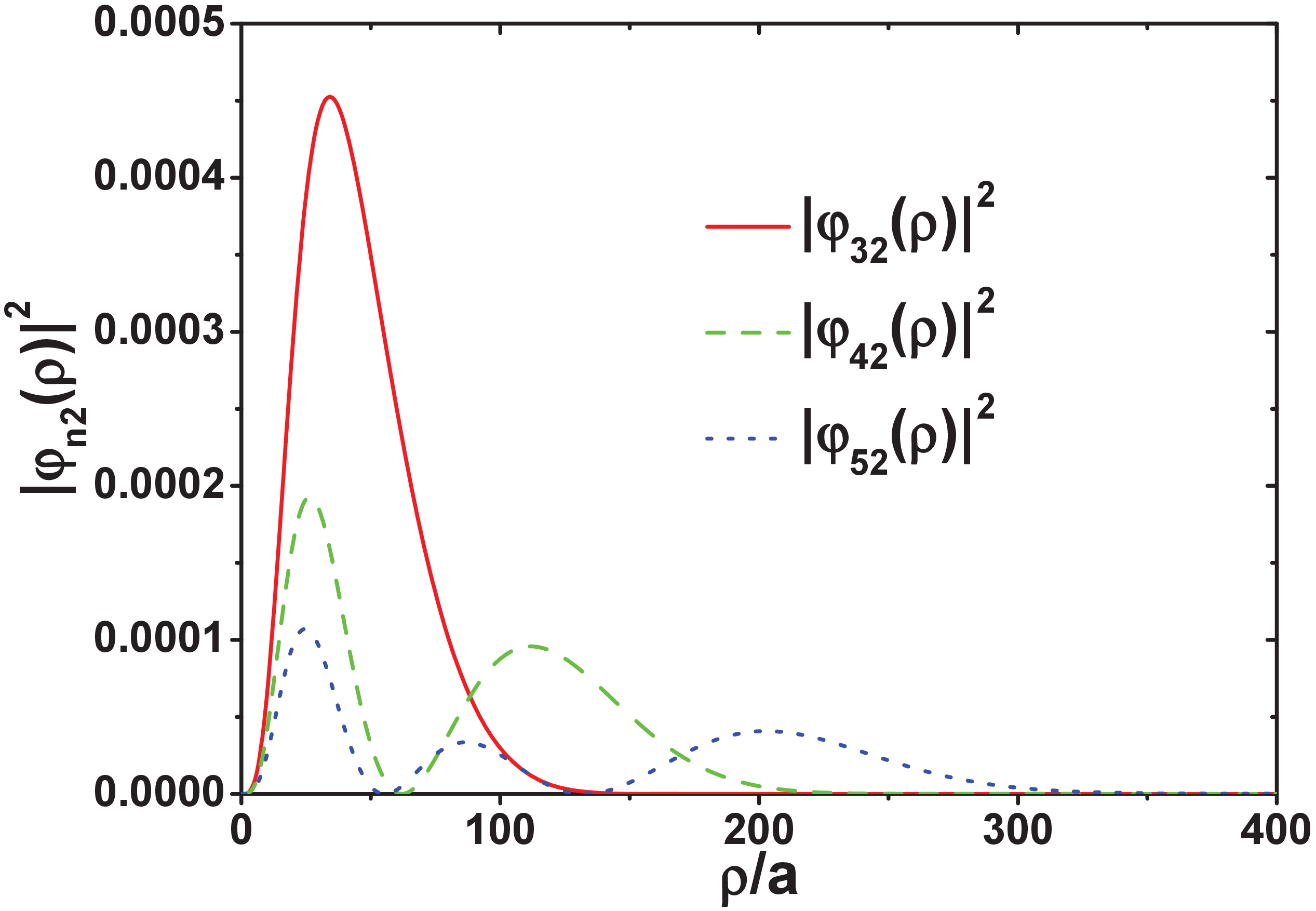} \\
\caption{Square modulus of the relative wave functions of the three lowest energy excitons, for (a) $m=0$, (b) $|m|=1$, (c) $|m|=2$. $t_{0}=-1$, $\Delta=0.2|t_{0}|$. The lattice constant $a$ is taken as the length unit. $r_{0}=\beta a=15a$. The inset of (a) is a magnified plot of the small magnitude part of $|\varphi_{10}(\rho)|^{2}$.}
\end{figure}

As shown in Fig. 4 are the square modulus of the relative wave function, $|\varphi_{nm}(\rho)|^{2}$, for several $\{n,m\}$ combinations. The parameters are the same as those for Fig. 3(a). For the same $|m|$, the wave function becomes more extended and develops more nodes as $n$ increases. As the quantum number $|m|$ increases, all the wave functions become increasingly more extended. Since electronic states in the flat band are spatially localized, the relative wave functions give mainly the distributions of the hole states of the dispersive valence band. The position of the electron states of the flat band defines the COM location of the exciton.

The approach we used to obtain the exciton spectrum and wave function, explained in Sec. IIC, makes it very convenient to make a comparison between the above results and the results for the corresponding 2D hydrogen problem defined by setting $\delta V(\rho)$ of Eq.(30) as zero. The energy spectrum of the corresponding 2D hydrogen is \cite{yang91}
\begin{equation}
E_{n}-|\Delta|=-\frac{1}{2(n-\frac{1}{2})^{2}}\frac{m^{\ast}e^{4}}{\hbar^{2}} =-\frac{1}{(n-\frac{1}{2})^{2}}\frac{|\Delta|}{3}(\frac{\beta}{\gamma})^{2},
\end{equation}
where $n=1,2,3,\cdots$. The lowest energy level, for $n=1$, is nondegenerate. For $n>1$, the energy level is $(2n-1)$-fold degenerate. In terms of the angular quantum number, these degenerate states correspond to $m=0, \pm1, \pm2, \cdots, \pm(n-1)$. Comparing to the results shown in Fig. 3(a) and 3(b), the deviation from the pure Coulomb interaction of Eq.(28) brings three major modifications to the spectrum. Firstly, the binding energies are reduced. From Eq.(34), $E_{1}-|\Delta|<-|\Delta|$ for all $\beta\ge \gamma$. However, the binding energy of the lowest exciton mode is smaller than the energy gap, as shown in Fig. 3(b). Secondly, the spacings between the consecutive energy levels shown in Fig. 3(a) do not follow the 2D hydrogenic model \cite{cudazzo11}. Thirdly, as is clear from Fig. 3(a), the degeneracy in the exciton levels for $n>1$ is broken. For the same $n$, the exciton level (exciton binding energy) decreases (increases) with increasing $|m|$.

The radial part of the excitons' relative wave functions are also changed qualitatively compared to the radial wave functions of the 2D hydrogen. On one hand, the $s$-like (i.e., $m=0$) exciton modes are more extended. In particular, instead of purely exponential decay, $|\varphi_{10}(\rho)|^{2}$ for the lowest exciton state has a local maximum away from $\rho=0$, as shown in Fig. 4(a) at around 11$a$. This results from the reduced attraction of the effective Coulomb interaction, which has the asymptotic behavior $V_{\text{eff}}(\rho)\sim -\ln(\rho)$ instead of $1/\rho$ for small $\rho$ \cite{cudazzo11}. Another finer change lies in the tails of the wave functions. For example, compared to the purely exponential decay of the wave function for the lowest level of the 2D hydrogen, there are additional oscillations in the tail of $|\varphi_{10}(\rho)|^{2}$, as shown in the inset of Fig. 4(a). The enhanced spatial extension of the exciton states is consistent with treating the intravalley excitons as large Wannier-Mott excitons for small $\Delta$.

Note that only the transverse spin-singlet $m=0$ excitons are optically active bright excitons and contribute to optical absorption for one-photon processes \cite{bassanibook}. The dark exciton states, which contain all the remaining exciton modes, may be probed by nonlinear two-photon processes via a virtual state \cite{bassanibook,ye14} or mediated by an intermediate state of the system \cite{heindel17}. In semiconductor quantum dots, manipulation of the dark exciton states in terms of single optical pulses are also possible, in the presence or absence of a magnetic field \cite{luker15,schwartz15}.


The intravalley excitons studied in this section differ from their counterparts in gapped monolayer graphene and the monolayer transition metal dichalcogenides in their completely localized COM motion. Consequently, the valley polarization generated by a circularly polarized light is spatially localized. This is peculiar in at least two aspects. Firstly, the valley polarized intravalley excitons are localized both in the real space and in the $\mathbf{k}$ space. This does not contradict the Heisenberg uncertainty principle because the localization in the real space is for the COM motion, while the localization in the $\mathbf{k}$ space corresponds to the spatially extended relative motion of the electron and hole states. Secondly, the spatially localized valley polarization associated with the intravalley excitons may allow for new applications in valleytronics and quantum information technology. For example, when the bright exciton generated by a circularly polarized light is turned to a dark exciton state through a mechanism such as those listed above, we get a long-lived valley-polarized state that can be used to store quantum information locally. The two intravalley exciton modes constitute a single qubit, which may be exploited to implement local quantum information processings through optical means.


The fact that the present intravalley excitons are dispersionless in their COM motion also makes these excitons stable to many perturbations that may change their COM momentum. This is because a process changing the COM momentum of the excitons must conserve both momentum and energy. We may consider changing the COM momentum of an exciton by coupling it to an internal (such as phonons) or external (such as photons) source of perturbation. The perturbation in general has a dispersion, and a finite momentum transfer accompanies a nonzero energy transfer. Since the dispersionless intravalley exciton has a flat spectrum, a usual perturbation with a dispersion cannot implement the momentum transfer because it would violate the energy conservation. As a result, the dispersionless excitons studied in this section should be stable to a large class of momentum transfer processes.


It is also interesting to compare with the core excitons, which appear in the core level spectra. In this case, an electron of a deep valence band, associated with a core atomic level, is optically excitated (e.g., by an X ray) to an unoccupied state of the conduction band. The core hole in the deep valence band may be considered as having an infinite mass, similar to the states in the flat band of the present system. As a result, the core excitons are completely localized in their COM motion, also similar to the intravalley excitons of the present system. On the other hand, the gap between the core hole level and the conduction band is much larger than the small gap considered here. Another more fundamental difference lies in the decay of the core hole by the Auger effect \cite{strinati84}. In consequence, it is of crucial importance to use dynamically screened interactions to make a realistic calculation for the core excitons, which are usually carried out by solving a Bethe-Salpeter equation \cite{strinati84,shirley98,wu15,qiu15}. In comparison, the hole states for the intravalley excitons of the symmetrically biased dice model considered in this work are at or close to the top of the valence band, and are not influenced by the Auger effect. As a result, a statically screened interaction like Eqs.(17) and (19) is reliable, and a treatment based on a hydrogenlike effective mass model is the standard formalism for Wannier-Mott excitons \cite{knoxbook,bassanibook,cudazzo11}.


\section{Effects of the exchange interaction}

The exchange interaction neglected in the previous discussions may introduce fine structures to the exciton spectrum. These include the splitting between the spin-singlet and spin-triplet excitons, and the splitting between the longitudinal and the transverse spin-singlet exciton modes \cite{onodera67,bassanibook}. For excitons of the symmetrically biased dice model, the exchange interaction also couples the intravalley excitons of the two valleys and therefore cause valley depolarization of the exciton states excitated by circularly polarized light, similar to the case of the monolayer transition metal dichalcogenides \cite{jones13}.

We firstly write down the definition of the exchange interaction as a term in the matrix element of the Coulomb interaction hybridizing two electron-hole pair states, where the electron state and the hole state of each pair belong to the same valley. As there is no spin-orbit coupling in the model, the total spin $M$ of the electron-hole pair is a good quantum number. The value of $M$ may be 0 (singlet states) or 1 (triplet states). The exchange interaction is nonzero only for singlet electron-hole pairs \cite{bassanibook,onodera67,pikus71}. Suppose $\Phi^{(M)}_{c\mathbf{k}_{e},v\mathbf{k}_{h}}$ is the many-body wave function (i.e., a Slater determinant) of the state with an electron of wave vector $\mathbf{k}_{h}$ annihilated from the fully occupied valence band and an electron of wave vector $\mathbf{k}_{e}$ created in the otherwise empty conduction band. The exchange interaction for the hybridization between $\Phi^{(M)}_{c\mathbf{k}_{e},v\mathbf{k}_{h}}$ and another state $\Phi^{(M)}_{c\mathbf{k}_{e}',v\mathbf{k}_{h}'}$ is \cite{bassanibook,onodera67}
\begin{eqnarray}
&& \hspace{-0.5cm} \langle\Phi^{(M)}_{c\mathbf{k}_{e},v\mathbf{k}_{h}}|\hat{H}|\Phi^{(M)}_{c\mathbf{k}_{e}',v\mathbf{k}_{h}'}\rangle_{\text{ex}}  \\ && \hspace{-0.5cm} =2\delta_{M}\langle\psi_{c\mathbf{k}_{e}}\psi_{v\mathbf{k}_{h}'}|\frac{e^{2}}{r_{12}}|\psi_{v\mathbf{k}_{h}}\psi_{c\mathbf{k}_{e}'}\rangle \notag \\
&& \hspace{-0.5cm} =2\delta_{M}\iint d\mathbf{r}_{1}d\mathbf{r}_{2} \psi_{c\mathbf{k}_{e}}^{\ast}(\mathbf{r}_{1})\psi_{v\mathbf{k}_{h}}(\mathbf{r}_{1}) \frac{e^{2}}{r_{12}}\psi_{c\mathbf{k}_{e}'}(\mathbf{r}_{2})\psi_{v\mathbf{k}_{h}'}^{\ast}(\mathbf{r}_{2}),  \notag
\end{eqnarray}
where $\psi_{c\mathbf{k}_{e}}$ ($\psi_{c\mathbf{k}_{e}'}$) and $\psi_{v\mathbf{k}_{h}}$($\psi_{v\mathbf{k}_{h}'}$) are separately the Bloch functions for an electron of wave vector $\mathbf{k}_{e}$ ($\mathbf{k}_{e}'$) in the conduction band and an electron of wave vector $\mathbf{k}_{h}$ ($\mathbf{k}_{h}'$) in the valence band. $\mathbf{k}_{e}-\mathbf{k}_{h}=\mathbf{k}_{e}'-\mathbf{k}_{h}'=\mathbf{k}_{\text{ex}}$ is the total wave vector of the electron-hole pairs. $\delta_{M}=1$ ($\delta_{M}=0$) for $M=0$ ($M=1$). When all the four wave vectors belong to the same valley $\mathbf{K}_{\tau}$, the exchange interaction couples electron-hole pairs of the same valley. We denote the intravalley exchange interaction as $J(\mathbf{k}_{\text{ex}})\delta_{M}$. When $\mathbf{k}_{e}$ and $\mathbf{k}_{h}$ are close to one valley $\mathbf{K}_{\tau}$ whereas $\mathbf{k}_{e}'$ and $\mathbf{k}_{h}'$ are close to the other valley $\mathbf{K}_{-\tau}$, the exchange interaction couples two electron-hole pairs of different valleys. We denote the intervalley exchange interaction as $\tilde{J}(\mathbf{k}_{\text{ex}})\delta_{M}$.

By expanding the Bloch functions in the above definition into a linear combination of Wannier functions, and retaining the leading order terms, it can be shown that both $J(\mathbf{k}_{\text{ex}})$ and $\tilde{J}(\mathbf{k}_{\text{ex}})$ have two parts, the long-range part and the short-range part \cite{bassanibook,onodera67,pikus71}. The short-range part is the same for $J(\mathbf{k}_{\text{ex}})$ and $\tilde{J}(\mathbf{k}_{\text{ex}})$, and is independent of $\mathbf{k}_{\text{ex}}$. The long-range part of the exchange interaction, also known as the nonanalytical part or the polarization term, depends on the COM momentum $\mathbf{k}_{\text{ex}}$ of the exciton \cite{bassanibook,onodera67,pikus71}.

Quantitative calculation of the short-range exchange interaction requires an accurate knowledge of the atomic orbitals on the lattice sites \cite{jones13,yu14b}. Because our discussions are based on a model rather than a concrete material, realistic evaluations of the short-range parts of $J(\mathbf{k}_{\text{ex}})$ and $\tilde{J}(\mathbf{k}_{\text{ex}})$ can not be made. The long-range exchange interaction for Wannier-Mott excitons of semiconductors, on the other hand, may be evaluated by the formula of Pikus and Bir in terms of the continuum approximation to the tight-binding model at the two valleys $\mathbf{K}_{\tau}$ ($\tau=\pm$) \cite{pikus71,yu14b,glazov14}. Direct application of the formula gives the following results for the intravalley and intervalley long-range exchange interactions
\begin{equation}
\begin{cases}
J_{\text{LR}}(\mathbf{k}_{\text{ex}})=\frac{3a^{2}t_{0}^{2}}{4\Delta^{2}}V(\mathbf{k}_{\text{ex}})k_{\text{ex}}^{2},    \\
\tilde{J}_{\text{LR}}(\mathbf{k}_{\text{ex}})=\frac{3a^{2}t_{0}^{2}}{4\Delta^{2}}V(\mathbf{k}_{\text{ex}})k_{\text{ex}}^{2} e^{2i\phi_{\tau}(\mathbf{k}_{\text{ex}})}.
\end{cases}
\end{equation}
We have multiplied Eq.(9) by the phase factor of Eq.(33) in calculating the eigenvector of the lower dispersive band at $\mathbf{K}_{\tau}$ ($\tau=\pm$). The phase $\phi_{\tau}(\mathbf{k})$ of a wave vector $\mathbf{k}$ is defined as $k\exp{i\phi_{\tau}(\mathbf{k})}=k_{x}+i\tau\text{sgn}(\Delta)k_{y}$. The bare 2D Coulomb interaction $V(\mathbf{k})=2\pi e^{2}/(\epsilon_{r}k)$. Both the intravalley and the intervalley long-range exchange interactions vanish linearly with $k_{\text{ex}}$ as $k_{\text{ex}}\rightarrow0$, which is well-known for 2D systems \cite{yu14,glazov14,yu14b,qiu15,andreani90}.


The four channels of the exchange interaction, intravalley versus intervalley and long-range versus short-range, introduce rich fine structures to the excitonic spectrum and dynamics. Firstly, since the exchange interaction is nonzero only for singlet excitons, there will be a singlet-triplet splitting (STS). Secondly, the intravalley long-range exchange interaction $J_{\text{LR}}(\mathbf{k}_{\text{ex}})$ gives a weak dispersion to the otherwise dispersionless intravalley exciton spectrum. Interestingly, this dispersion is linear and isotropic in $\mathbf{k}_{\text{ex}}$. On one hand this implies the absence of longitudinal-transverse splitting (LTS) by $J_{\text{LR}}(\mathbf{k}_{\text{ex}})$ in the intravalley excitons. On the other hand the intravalley excitons gain mobility through $J_{\text{LR}}(\mathbf{k}_{\text{ex}})$. So, the prospective applications mentioned at the end of the last section apply only when the mobility from $J_{\text{LR}}(\mathbf{k}_{\text{ex}})$ is sufficiently small. Noth that, the dispersion induced by $J_{\text{LR}}(\mathbf{k}_{\text{ex}})$ corresponds to the nonanalyticity in the dispersion of the excitons of monolayer MoS$_{2}$ found by Qiu et al \cite{qiu15}. Thirdly, the \emph{intervalley} exchange interactions $\tilde{J}(\mathbf{k}_{\text{ex}})$ mediate nontrivial coupling between the intravalley excitons of the two valleys. On one hand, for the intravalley excitons generated by circularly polarized light, the intervalley exchange interaction acts as a mechanism of valley depolarization \cite{yu14b}. On the other hand, the intervalley exchange interaction brings hybridization and level splitting to the two intravalley exciton modes, when the system is excitated by a linearly polarized light which excitates the two intravalley excitons simultaneously in equal strength. As was demonstrated in the corresponding case of monolayer transition-metal dichalcogenides, the two levels resulting from this splitting are separately longitudinal and transverse combinations of the two intravalley excitons \cite{yu14,glazov14,wang18}. Therefore, the intervalley exchange interactions lead to a longitudinal-transverse splitting in the hybridized non-valley-polarized exciton modes. Only the transverse singlet excitons can be excitated by light and may be bright excitons \cite{bassanibook}. In addition, this hybridization and splitting of the two degenerate intravalley exciton modes tends to establish a coherence between the two valleys, such that the photoluminescence following the excitation retains the linear polarization of the stimulus light pulse \cite{jones13}.


Now, by taking $J$ as a free parameter, we make a quantitative calculation of the variation in the intravalley exciton spectrum induced by the intravalley exchange interaction. For the intravalley singlet excitons, the intravalley exchange interaction adds the following term to the effective-mass model of Eq.(16),
\begin{equation}
J\delta(\mathbf{r}),
\end{equation}
where $\delta(\mathbf{r})=\delta(x)\delta(y)$ is the 2D Dirac delta function \cite{knoxbook,bassanibook,onodera67}. If we are interested only in the $\mathbf{k}_{\text{ex}}=0$ excitons, the parameter $J$ contains only the short-range component. The Dirac delta function implies that the exchange term has a very mild impact on the spatially extended Wannier-Mott excitons, which somehow justifies the omission of the exchange term in studying large Wannier-Mott excitons \cite{cudazzo11}. To incorporate Eq.(37) into the formalism of Sec.IIC, we approximate the Dirac delta function as the limit of a Gaussian,
\begin{equation}
\delta(\mathbf{r})=\delta(x)\delta(y)=\lim_{\sigma\rightarrow0}\frac{1}{2\pi\sigma}e^{-\frac{x^{2}+y^{2}}{2\sigma}} =\lim_{\sigma\rightarrow0}\frac{1}{2\pi\sigma}e^{-\frac{\rho^{2}}{2\sigma}}.
\end{equation}
In this manner, for each definite $\sigma$$>$$0$ the exchange interaction depends only on the polar radius $\rho$ and can be directly added to the residual potential of Eq.(30). For a specific value of $J$, we calculate the modified spectrum of the intravalley excitons for successively decreasing and positive $\sigma$, until the lowest eigenenergies attain convergence. We therefore test simultaneously the convergence of the spectrum with respect to the number $N$ of retained basis and the parameter $\sigma$, for sufficiently large $L$ for the integration of Eq.(31).

\begin{figure}[!htb]\label{fig5} \centering
\includegraphics[width=8.5cm,height=6.669cm]{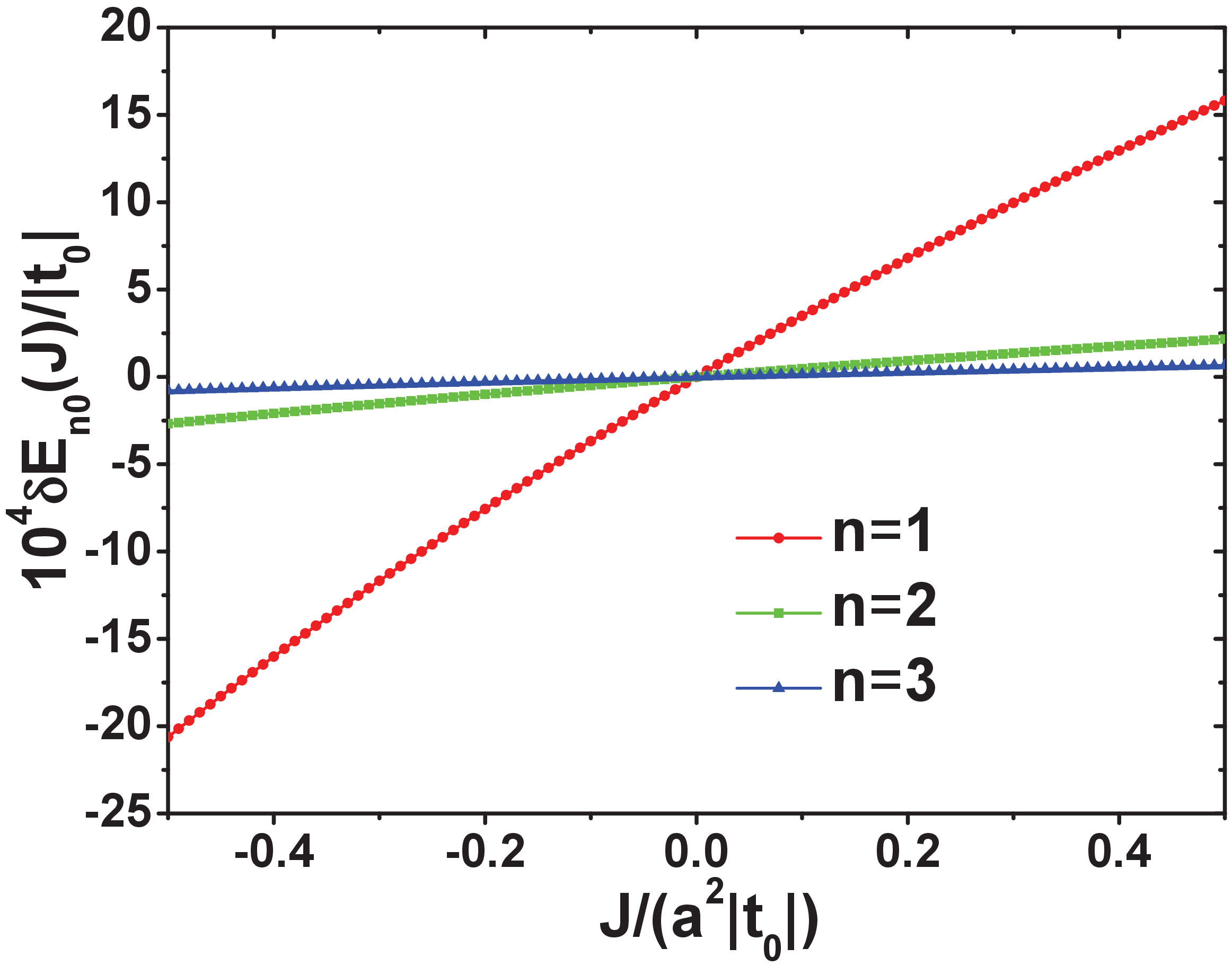}
\caption{Dependency of the three lowest $s$-wave (i.e., $m=0$) excitons on $J$. What is plotted is the variation of the energies, $\delta E_{n0}(J)=E_{n0}(J)- E_{n0}(J=0)$, in unit of $|t_{0}|$. The Dirac delta function is approximated by Eq.(38), with $\sigma=10^{-6}a^{2}$. The other parameters are the same as those for Fig. 3(a) and Fig. 4.}
\end{figure}

We have made extensive numerical calculations for the modified spectrum of singlet intravalley excitons by incorporating the exchange interaction in the above manner. One central conclusion is that only the $s$-wave (i.e., $m=0$) intravalley excitons are slightly changed after including the intravalley exchange interaction. This is consistent with the local nature of the exchange interaction and the radial distribution of the relative wave functions shown in Fig. 4, from which only the $s$-wave excitons have nonzero value at $\rho=0$ (i.e., at $\mathbf{r}=0$). The convergence in the exciton energy levels is attained when $\sigma\le10^{-3}a^{2}$. As shown in Fig. 5 are the variations of the three lowest $s$-wave excitons with $J$, evaluated at $\sigma=10^{-6}a^{2}$ to ensure the convergence of the results. From Fig. 5, and in comparison to Fig. 4(a), the variation of the energy levels $\delta E_{n0}(J)=E_{n0}(J)- E_{n0}(J=0)$ is roughly proportional to the product $J|\varphi_{n0}(0)|^{2}/a^{2}$, in particular for small $J$. This is in agreement with the above picture for why the excitons with $|m|\ge1$ are not affected by the exchange term.

Because the exchange interaction acts only on the spin-singlet excitons, the results in Fig. 5 imply a singlet-triplet splitting (STS) for the $s$-wave excitons ($m=0$). The STS is absent for $|m|>0$. For $J>0$, the triplet $s$-wave excitons, which are dark excitons, have lower energy and are the lowest energy exciton excitations. For $J<0$, the singlet $s$-wave excitons, which are bright excitons (the transverse branch), have lower energy and are instead the lowest energy exciton excitations. Since there is a sign reversal in the exchange interaction compared to the attractive direct Coulomb interaction \cite{bassanibook}, the parameter $J$ is expected to be positive in usual cases.

\section{summary}

In summary, we have studied the interband transitions and excitons in the symmetrically biased dice model. We find valley-contrasting interband transitions between the flat band and one dispersive band, under excitations by circularly polarized light. In terms of the static polarizability obtained by RPA calculations, the excitons in the system may be regarded as Wannier-Mott excitons when the bias is small compared with the hopping amplitude of the dice model. Based on the effective-mass model for the intravalley excitations, we calculate the spectrum and the wave functions of the intravalley excitons. Fine structures in the exciton spectrum, induced by the exchange interaction, are also discussed. The present study indicates that the symmetrically biased dice model is a new platform for studying valley-contrasting optoelectronics.



\end{document}